\documentclass[aps,pre,twocolumn,floatfix,superscriptaddress]{revtex4-1}
\usepackage{graphicx}
\usepackage{amsmath,amssymb,latexsym}
\usepackage{amsfonts}
\usepackage{color}
\usepackage{epsfig}
\usepackage{multirow}
\usepackage{floatrow}
\usepackage{tabu}
\usepackage{epstopdf}

\def\Dm{{\Delta m}}

\begin{document}

\title{Self-similar aftershock rates}
\author{J\"{o}rn Davidsen}
\email[]{davidsen@phas.ucalgary.ca}
\affiliation{Complexity Science Group, Department of Physics and Astronomy, University of Calgary, Canada}
\affiliation{GFZ German Research Centre for Geosciences, Section 3.2: Geomechanics and Rheology, Telegrafenberg, D14473 Potsdam, Germany}
\author{Marco Baiesi}
\email[]{baiesi@pd.infn.it}
\affiliation{Department of Physics and Astronomy, University of Padova, Via Marzolo 8, I-35131 Padova, Italy}
\affiliation{INFN - Sezione di Padova, Via Marzolo 8, I-35131 Padova, Italy}

\date{\today}

\begin{abstract}
In many important systems exhibiting crackling noise --- intermittent avalanche-like relaxation response with power-law and, thus, self-similar distributed event sizes --- the ``laws'' for the rate of activity after large events are not consistent with the overall self-similar behavior expected on theoretical grounds. This is in particular true for the case of seismicity and a satisfying solution to this paradox has remained outstanding. Here, we propose a generalized description of the aftershock rates which is both self-similar and consistent with all other known self-similar features. Comparing our theoretical predictions with high resolution earthquake data from Southern California we find excellent agreement, providing in particular clear evidence for a unified description of aftershocks and foreshocks. This may offer an improved way of time-dependent seismic hazard assessment and earthquake forecasting. 
\end{abstract}
\maketitle

\section{Introduction}

Many natural and man-made systems exhibit an intermittent avalanche-like response to changing external conditions~\cite{bak,turcotte99}. Sequences of such sudden responses or events often constitute the most crucial features of the evolutionary dynamics of complex systems, both in terms of their description, characterization and understanding. Prominent examples include earthquakes on fault systems~\cite{benzion08}, frictional sliding~\cite{goebel13}, irreversible plastic deformation in solids~\cite{miguel01,weiss04,csikor07}, fracture~\cite{bonamy11,tantot13,baro13,kun14,maekinen15,ribeiro15}, materials failure~\cite{zapperi97,negri15}, magnetization processes~\cite{durin06,papanikolaou11}, solar flare emissions~\cite{baiesi06m,arcangelis06a}, financial markets~\cite{lillo03,weber07,petersen10,siokis12}, internet traffic~\cite{abe03a}, and media coverage~\cite{klimek11}. The avalanche-like response --- often called crackling noise --- is characterized by discrete, impulsive events spanning a broad range of energies $E$, with a power-law frequency distribution $P(E)\propto E^{-\epsilon}$ ~\cite{sethna01,laurson13,salje14,uhl15}. The empirical Gutenberg-Richter (GR) relation for earthquakes is one specific example~\cite{gutenberg}: Energies released by earthquakes follow a power-law distribution and are conveniently handled in the logarithmic scale of the magnitude $m$ with $E \propto 10^{\frac{3}{2}m}$ such that $P(m)\propto 10^{-b m}$, where $b = \frac{3}{2} (\epsilon - 1)$. There is also a good degree of universality, as $\epsilon$ is close to $1.5$ for many systems exhibiting crackling noise~\cite{uhl15}. The associated absence of characteristic scales indicates scale free or self-similar behavior. 

Significant progress has been made in understanding the self-similar distribution of event sizes and the associated universality of crackling noise by using mean-field and renormalization approaches~\cite{sethna01,corral05,zaiser06,doussal09,papanikolaou11,dahmen11,laurson13,salje14,uhl15}. These theoretical approaches elevate self-similar behavior to a general principle as in the case of critical phenomena in equilibrium systems. Thus, it is important to establish whether other properties often associated with crackling noise also exhibit self-similar behavior. This includes those spatio-temporal correlations between events that reflect the intrinsic or endogenous dynamics of a given system and are a consequence of event-event triggering as for aftershocks~\cite{weiss04,crane08,sornette09,gu13amj,hainzl14,stojanova14}. This is most clearly reflected in the time-varying (local) event rates following large events, which are empirically found to follow --- across a wide range of scales and systems from friction and fracture to socio-economic systems~\cite{utsu95,gu13amj,moradpour14,davidsen14,goebel13a,baro13,maekinen15,ribeiro15,arcangelis06a,lillo03,weber07,petersen10,siokis12,abe03a,klimek11} --- the Omori-Utsu (OU) relation,
\begin{equation}
r(t) = \frac{K}{(t + c)^p} \equiv \frac{1}{\tau (t / c + 1)^p },
\label{eq:ou}
\end{equation}
first proposed for earthquakes~\cite{omori}. Here, $t$ measures the time after the large event or trigger, $p$ is typically close to 1 ($p\gtrsim 1$ if one only considers directly triggered events~\cite{davidsen14}) and $\tau \equiv {c^p}/{K}$. $K$ is found to increase with the energy of the trigger, according to the {\em productivity} relation $K = K_0 E^{2\alpha/3}$. Its equivalent formulation in terms of the magnitude of the trigger, $M$, is $K = K_0 10^{\alpha M}$. The exponent $\alpha$ is less than $b$ across many systems exhibiting crackling noise~\cite{weiss04,hainzl08,petersen10,gu13amj,baro13,ribeiro15,maekinen15}. 
The parameter $K_0$ naturally depends on the observational threshold $m_{\rm{th}}$~\cite{corral03,corral05}, by lowering it one counts more triggered events and it was reported  
$K_0 \sim 10^{-\beta m_{\rm{th}}}$~\cite{shcherbakov06,nanjo07,bhattacharya11,davidsen14}. In principle the exponents $\alpha$, $\beta$, and the $b$-value for triggered events or aftershocks~\cite{gu13amj,shearer12}, $b_{\rm{as}}$, appearing in the mentioned scalings may be different and indeed this is often observed~\cite{helmstetter03a,hainzl08,chu11,zaliapin13a,gu13amj,davidsen14}.

A consequence of the difference between $\alpha$ and $b_{\rm{as}}$ across many systems exhibiting crackling noise is the breakdown of self-similarity in the triggering process in those cases~\cite{vere-jones05}. Specifically, the number of triggered events of a given energy will explicitly depend on the energy of the trigger and not just their energy ratio: Since the magnitudes of the triggered events or aftershocks are distributed according to $P(m_{\rm{as}})\propto 10^{-b_{\rm{as}} m_{\rm{as}}}$, the number of events with magnitude $m_{\rm{as}}$ triggered by an event of magnitude $M$ scales as $10^{\alpha M - b_{\rm{as}} m_{\rm{as}}}$ for constant $c$ in Eq.~\eqref{eq:ou}. This translates into a scaling with $10^{(\alpha-b_{\rm{as}})M + b_{\rm{as}} \Delta m}$ with $\Delta m = M - m_{\rm{as}}$ that includes an explicit dependence on $M$ for $\alpha \neq b_{\rm{as}}$, in disagreement with the hypothesis of a self-similar triggering process that only depends on $\Delta m$~\cite{corral03,corral05,shcherbakov15,vere-jones05,saichev05a,turcotte07,lippiello07,lippiello07a}. An explicit dependence on $M$​ would imply that the triggering process inducing a number of events of, say, magnitude $6$​ due to an event of magnitude $8$​ is fundamentally different from the triggering process inducing a number of events of magnitude $3$ due to an event of magnitude $5$.

To reconcile different values of $\alpha$ and $b_{\rm{as}}$ with self-similar triggering, we build on the well-established behavior of equilibrium critical phenomena and propose a natural generalization of the OU relation that is consistent with a self-similar triggering process. Specifically, we build on the fact that critical phenomena can be characterized not only by critical exponents but also by universal scaling functions that describe the behavior near equilibrium critical points~\cite{christensen}. A general way to cast the event-event triggering rates into such a scaling form under the condition that the rates should only depend on the energy ratios between trigger and triggered event or, equivalently, their magnitude difference $\Delta m = M - m_{\rm{as}}$ is the following ansatz
\begin{equation}\label{eq:f}
	r(m_{\rm{as}},t|M,0)=\dfrac{1}{\tau_{\Dm}} f\left(\frac t {c_{\Dm}}\right),
\end{equation}
where $\tau_\Dm$ and $c_\Dm$ are two time scales varying only with $\Delta m$. In fact, an approach based on a limited scaling form with a constant $\tau_\Dm = \tau$ was previously introduced by Lippiello \emph{et al.}~\cite{lippiello07,lippiello07a}. In this paper, by analyzing high-resolution earthquake data from Southern California, we show that only the full self-similar generalization of the OU relation, in a form following Eq.~\eqref{eq:f} with both $c_{\Dm}$ and $\tau_{\Dm}$ scaling with $\Delta m$ and with a specific functional form of $f$, is consistent with all empirical relations. In this generalized form of the OU relation, self-similarity is present even in the case of a non-constant $c$ and $\alpha \neq b_{\rm{as}}$.

\begin{figure*}[!t]
\begin{center}  
\begin{minipage}{0.66\textwidth}
  \centering
  \includegraphics[width = 0.99\textwidth]{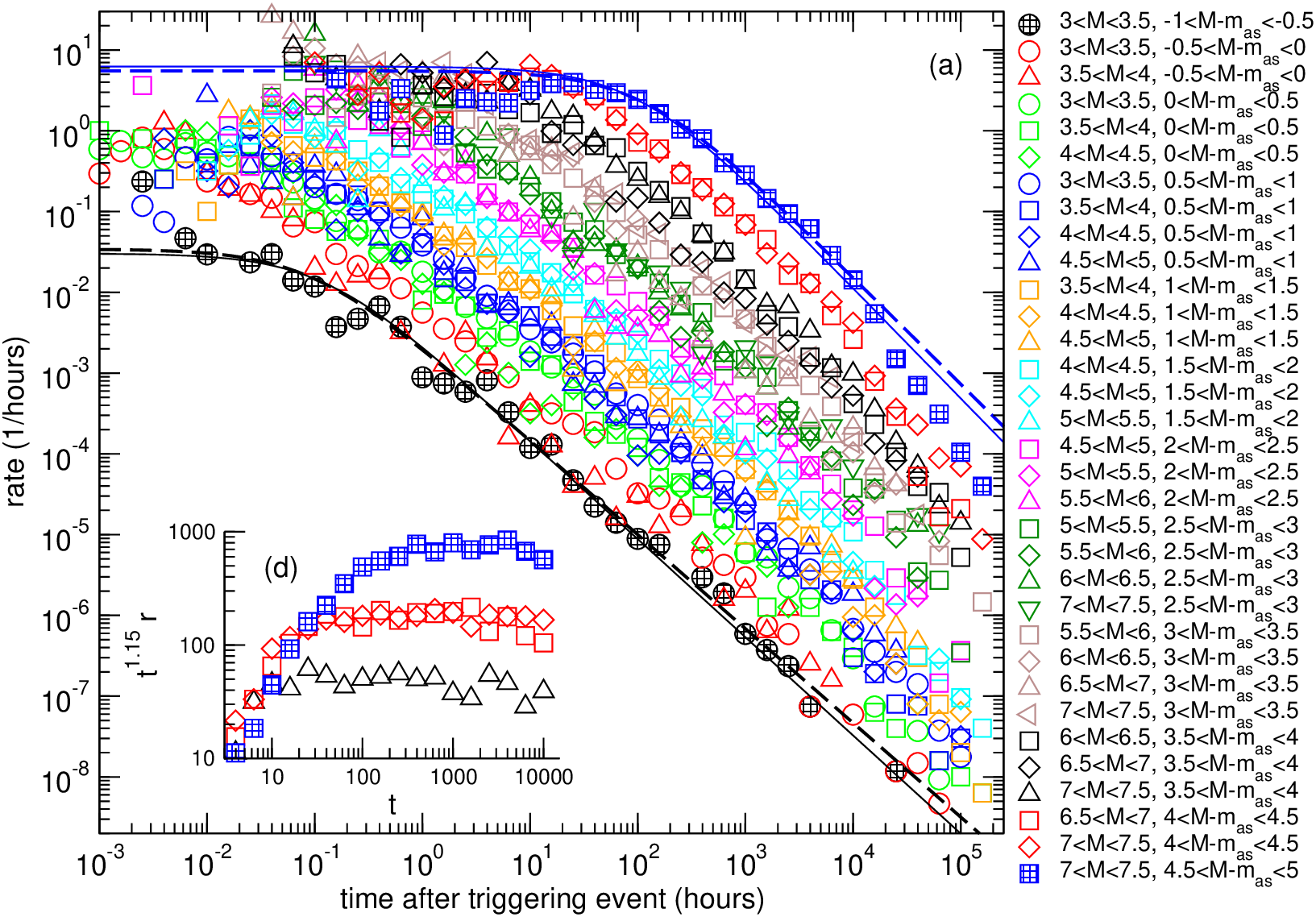}
\end{minipage}\hfill
\begin{minipage}{0.31\textwidth}
  \centering
  \includegraphics[width = 0.99\textwidth]{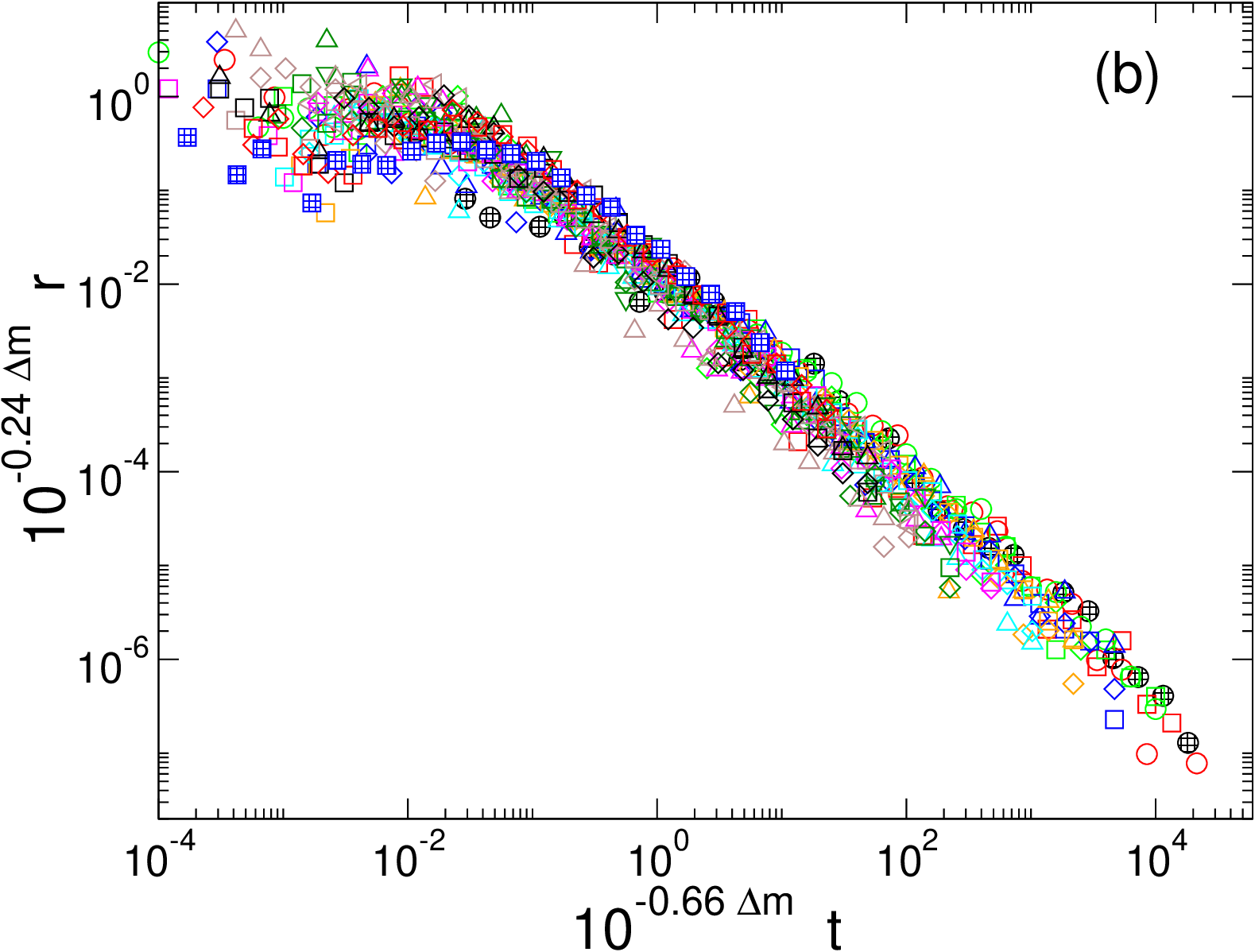}\\
\vskip 2mm
  \includegraphics[width = 0.99\textwidth]{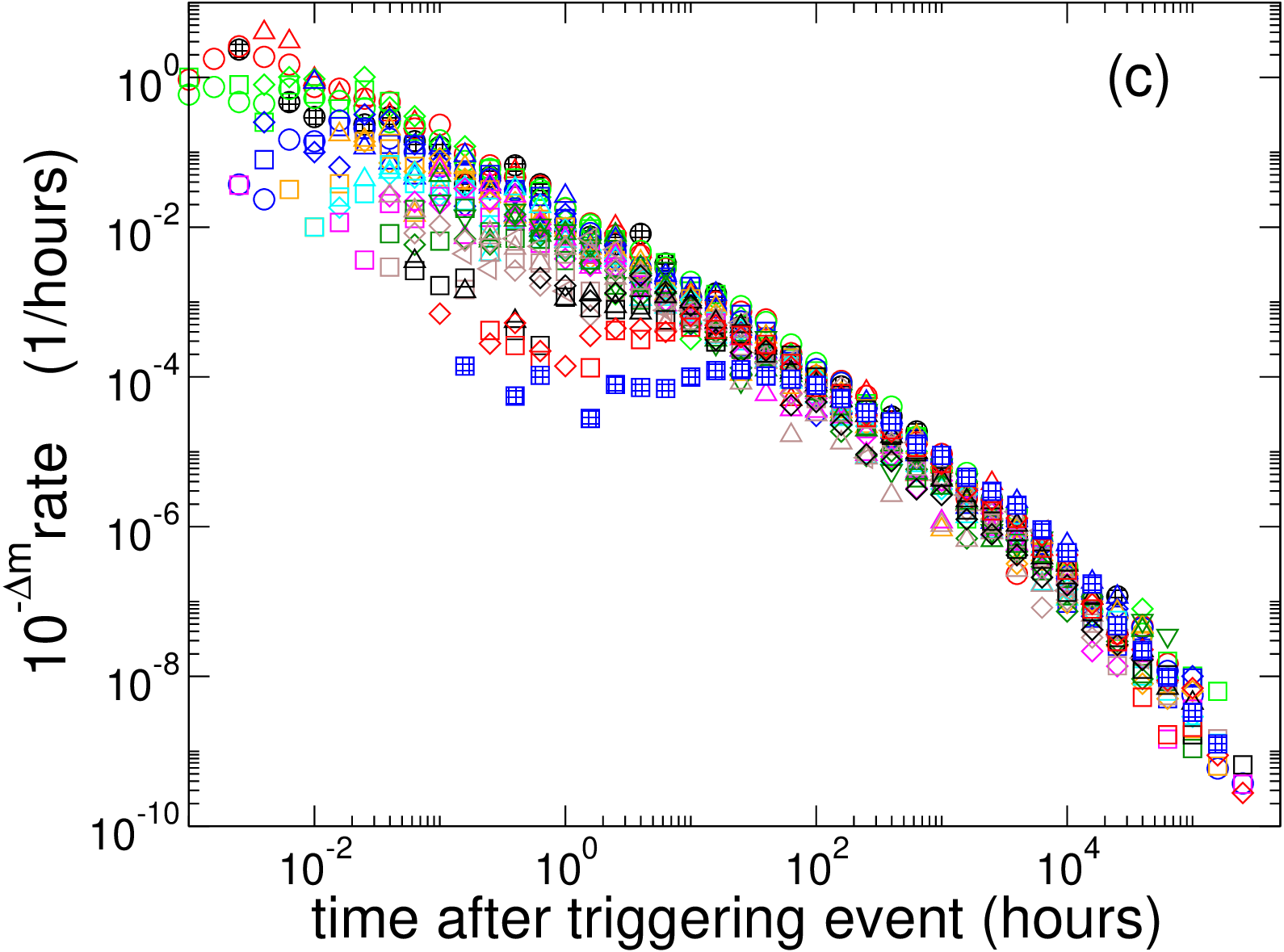}
\end{minipage}\hfill
\caption{\label{fig:ou}(a) Averaged conditional aftershock rates for different ranges of main shock and aftershock magnitudes. Note that all curves with the same magnitude difference collapse. The solid lines correspond to OU fits over the full range of $t$, while the dashed lines correspond to fits up to $t=10^4$ hours only. See Fig.~\ref{fig:fit} for all the estimated OU parameters. (b) All curves collapse onto a unique scaling function under appropriately chosen rescaling with the magnitude difference as predicted by the scaling ansatz~\eqref{eq:ss_ou}. Data with $t>10^4$ hour are not considered here due to the natural detection problem discussed in the main text.
(c) Conditional rates rescaled by $10^{-\alpha \Dm}$ with $\alpha = 1$. (d) Some conditional rates rescaled by $t^{p}$ with $p = 1.15$.
}
\end{center}
\end{figure*}

\section{Results}
\subsection{Generalized OU relation \label{sec:ou}}

Self-similarity in the OU relation~\eqref{eq:ou} is violated if the \emph{number} of the directly triggered events of magnitude $m_{\rm{as}}$ does explicitly depend on the main shock magnitude $M$, and not only on the magnitude difference. To ensure self-similarity and consistency with empirical observations, we propose the following generalization of the OU relation for \emph{conditional} event-event triggering rates:
\begin{equation}\label{eq:ss_ou}
	r(m_{\rm{as}},t|M,0)=\dfrac{1}{\tau_{\Delta m}} \left( \dfrac{t}{c_{\Delta m}} + 1 \right)^{-p},
\end{equation}
with time scales
\begin{equation}\label{eq:c_tau}
c_{\Dm}=c_0 10^{ g\,  \Dm} \quad \textrm{and} \quad \tau_{\Dm}=\tau_0 10^{- z\,  \Dm}
\end{equation} 
scaling with $\Dm$ ($ g \geq 0 $ and $ z \geq 0$ are supposedly universal scaling exponents while $c_0$ and $\tau_0$ are constant prefactors). Specifically, $r(m_{\rm{as}},t|M,0)$ is the rate of events or aftershocks of magnitude $m_{\rm{as}}$ at time $t$ triggered directly by an event of magnitude $M$ at time $0$. This generalized OU relation corresponds to Eq.~\eqref{eq:f} with $f(x) = (1+x)^{-p}$, hence it is a natural generalization in the sense that $f$ is consistent with the standard OU form and all parameters now depend explicitly and exclusively on $\Dm$. Note also that Eq.~\eqref{eq:ss_ou} does not require $M>m_{\rm{as}}$ but it is applicable to all magnitude or energy combinations such that it encompasses what is often considered main shock-aftershock pairs as well as foreshock-main shock pairs. 

With a simple mathematical derivation (see \cite{shcherbakov15} for a somewhat similar derivation in a context in which the magnitude of the largest aftershock is assumed to play a significant role),
we may show that Eq.~\eqref{eq:ss_ou} ensures self-similarity, as the total number of triggered events of magnitude $m_{\rm{as}}$,
\begin{equation}\label{eq:ss_n}
	N(m_{\rm{as}}|M) \equiv \int_0^\infty  r(m_{\rm{as}},t|M,0)\, d t = \dfrac{c_{\Delta m}}{\tau_{\Delta m} (p-1)},
\end{equation}
depends only on $\Dm$.	 
Such self-similar generalization of the OU relation is consistent with the GR relation since
\begin{eqnarray}\label{eq:ss_gr}
	N_>(m_{\rm{th}}|M) &\equiv& \int_{m_{\rm{th}}}^{\infty} N(m_{\rm{as}}|M)\, d m_{\rm{as}}\\
                          &=&      \dfrac{c_0}{\tau_0 (p-1) \ln{10} (g+z)} 10^{(g+z)(M-m_{\rm{th}})} . \nonumber
\end{eqnarray}
Thus, we have the scaling relation
\begin{equation}\label{eq:sr_1}
b_{\rm{as}} = g + z,
\end{equation}
indicating that only two out of these three scaling exponents are independent. Hence, the generalized OU relation does not introduce an additional independent parameter compared to the standard OU relation with its associated productivity relation. In particular, for $K_{\Dm} \equiv {c_{\Dm}^p}/{\tau_{\Dm}}$ we also have a generalized productivity relation 
\begin{equation}\label{eq:prod}
K_{\Dm}=K_0 10^{\alpha \Dm}
\end{equation} 
with $K_0=c_o^p/\tau_0$ and 
\begin{equation}\label{eq:sr_2}
\alpha= z  + p  g =b_{\rm{as}} +  g  (p-1).
\end{equation} 
Note that this implies that $K_{\Dm}$ and $N(m_{\rm{as}}|M)$ do \emph{not} scale the same way with $\Dm$ if $ g  \neq 0$, which explicitly allows $b_{\rm{as}} \neq \alpha$ in our self-similar framework.

A related consequence of the generalized OU relation given by Eqs.~\eqref{eq:ss_ou} and~\eqref{eq:c_tau} is that the GR relation for triggered events needs to be modified if only triggered events over short time intervals are considered. For example, the number of triggered events or aftershocks of magnitude $m_{\rm{as}}$ up to time $t^*$, 
\begin{equation}\label{eq:ss_n_tr}
	N(m_{\rm{as}},t^*|M) \equiv \int_0^{t^*} r(m_{\rm{as}},t|M,0)\, d t,
\end{equation}
only follows the GR relation~\eqref{eq:ss_n} for $t^*\to\infty$. For finite $t^*$, $N(m_{\rm{as}},t^*|M)$ has two power-law regimes: For small $m_{\rm{as}}$, it decays with exponent $z$ while it decays with exponent $b_{\rm as}=g+z$ for large $m_{\rm{as}}$. The transition point is around a magnitude $m^*$ for which $c_{M-m^*}\approx t^*$. Thus, the transition point moves to lower magnitudes for increasing $t^*$, recovering the full GR relation for $t^*\to\infty$. Note that only for $g=0$ --- corresponding to a constant $c_{\Dm}$ --- the GR relation holds for all $t^*$.

In contrast to the conditional rates in Eqs.~\eqref{eq:f} and~\eqref{eq:ss_ou}, the classic OU relation~\eqref{eq:ou} considers the integrated rates $r_>(m_{\rm{th}},t|M,0) \equiv \int_{m_{\rm{th}}}^\infty r(m_{\rm{as}},t|M,0)\, d m_{\rm{as}}$. For the proposed self-similar OU conditional rates in Eq.~\eqref{eq:ss_ou}, one easily verifies
that such integrated rates inherit the scale invariance with respect to $M-m_{\rm{th}}$. Moreover, these integrated rates have a functional form very similar to the classic OU relation for realistic situations. 
See the Supplementary Information for more details. 

To summarize, the proposed self-similar form of the conditional event-event triggering rates can indeed be considered a realistic generalization of the classic OU relation as it is consistent with the GR relation, the productivity relation as well as the classic form of the OU relation for integrated rates.

\subsection{Comparison with data \label{sec:data}}

To test the validity of the self-similar OU relation, we analyze the event-event triggering for earthquakes in Southern California. The triggering relations between earthquakes are identified using the established methodology described in {\em Methods}.

\paragraph*{Self-similarity of conditional rates:}
 In Fig.~\ref{fig:ou}(a) we show the conditional triggering rates for different combinations of magnitudes. The striking feature is that all rates with a given $\Dm$ are quite indistinguishable from each other, regardless of the magnitude of the trigger. This strongly suggests that $\Dm$ is the relevant quantity determining the triggering rates and, thus, supports the hypothesis of self-similarity as formulated in Eq.\eqref{eq:f}. Since this behavior is independent on whether $\Dm$ is positive or negative (see Fig.~\ref{fig:ou}(a)), it also provides a unified description of aftershocks and foreshocks. 

To further establish that the dependence on $\Dm$ is correctly captured by Eq.~\eqref{eq:c_tau}, we recall that Eq.~\eqref{eq:ss_ou} is an example of the general scaling form~\eqref{eq:f}.	
This implies that all curves should collapse onto a single master curve, the scaling function $f(x)$, under appropriate rescaling with $c_{\Dm}$ and $\tau_{\Dm}$. This is indeed what we observe in Fig.~\ref{fig:ou}(b) for $g \approx 0.66$ and $z \approx 0.24$ providing direct support for the scaling proposed in Eq.~\eqref{eq:c_tau}.

\begin{figure}[!tb]
\begin{center}
\includegraphics[width = 0.46\columnwidth]{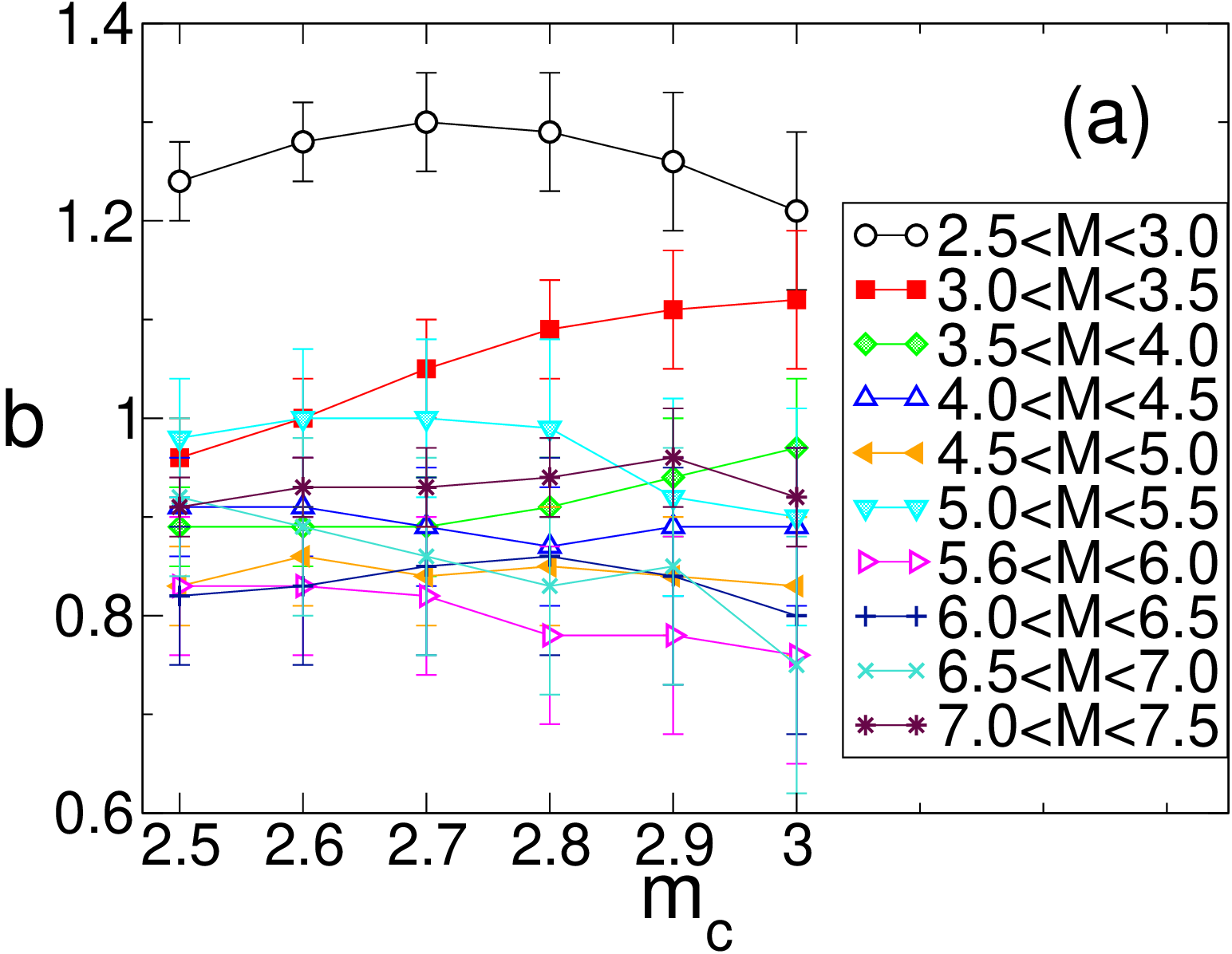}
\hskip 1mm
\includegraphics[width = 0.46\columnwidth]{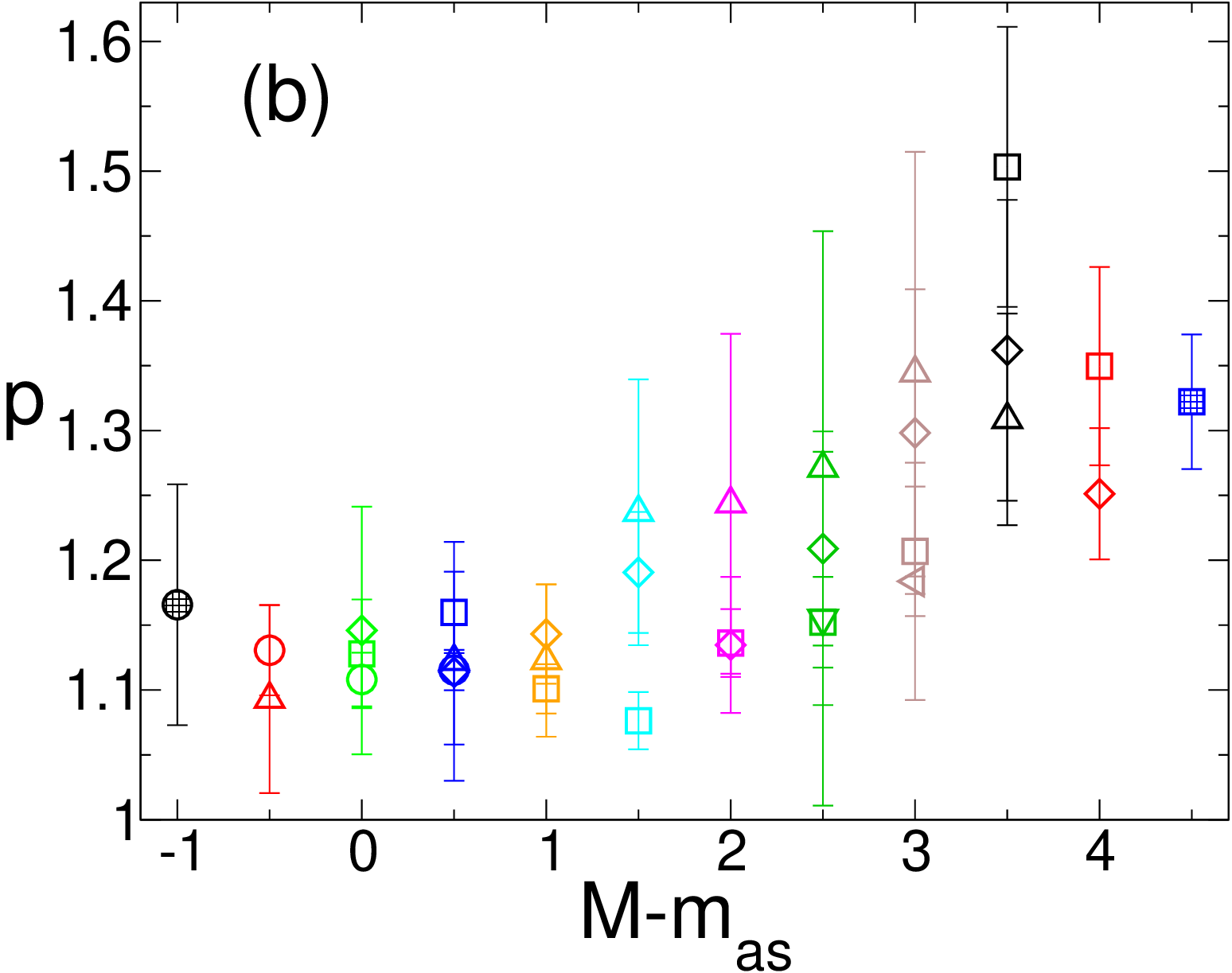}
\vskip 1mm
\includegraphics[width = 0.46\columnwidth]{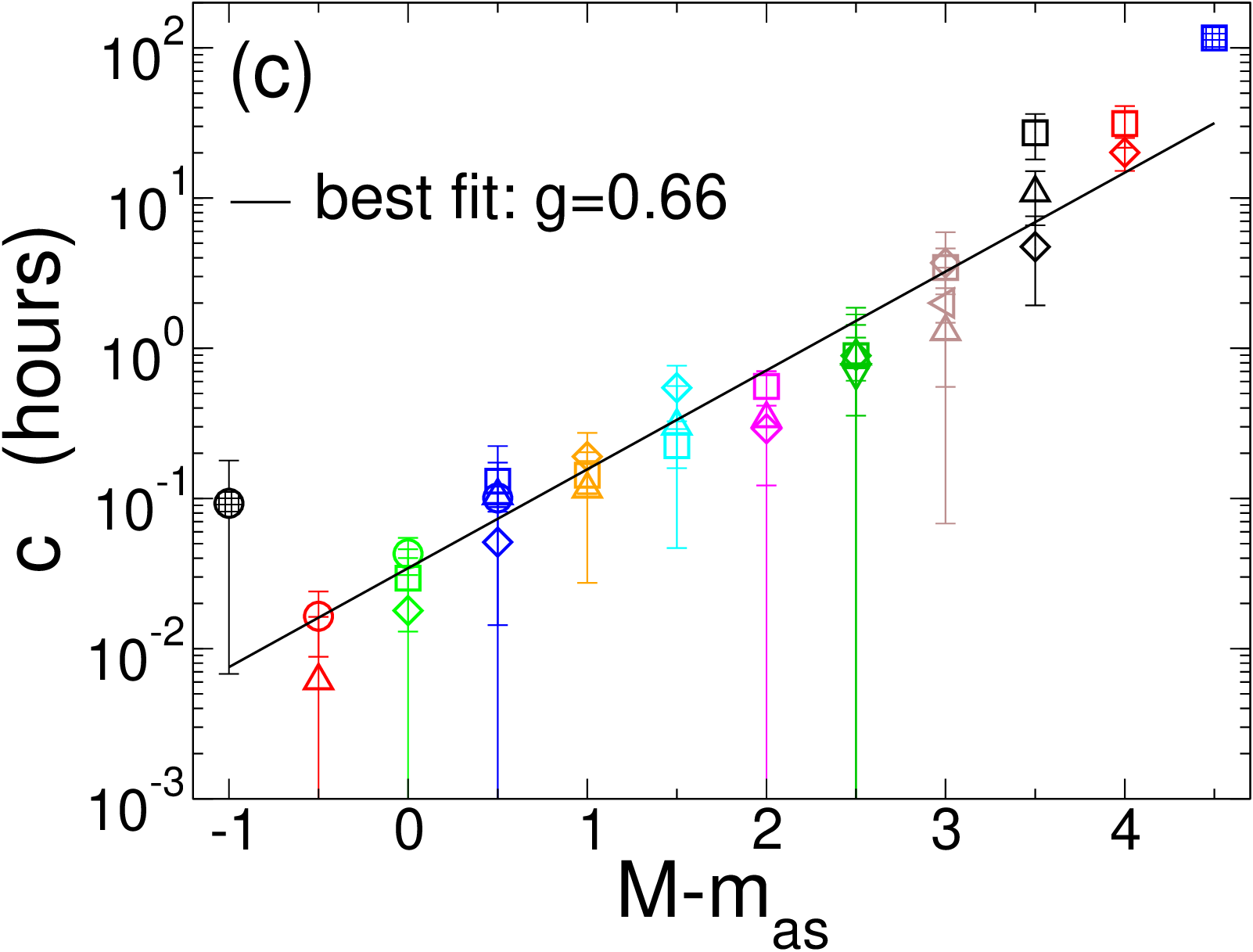}
\hskip 1mm
\includegraphics[width = 0.46\columnwidth]{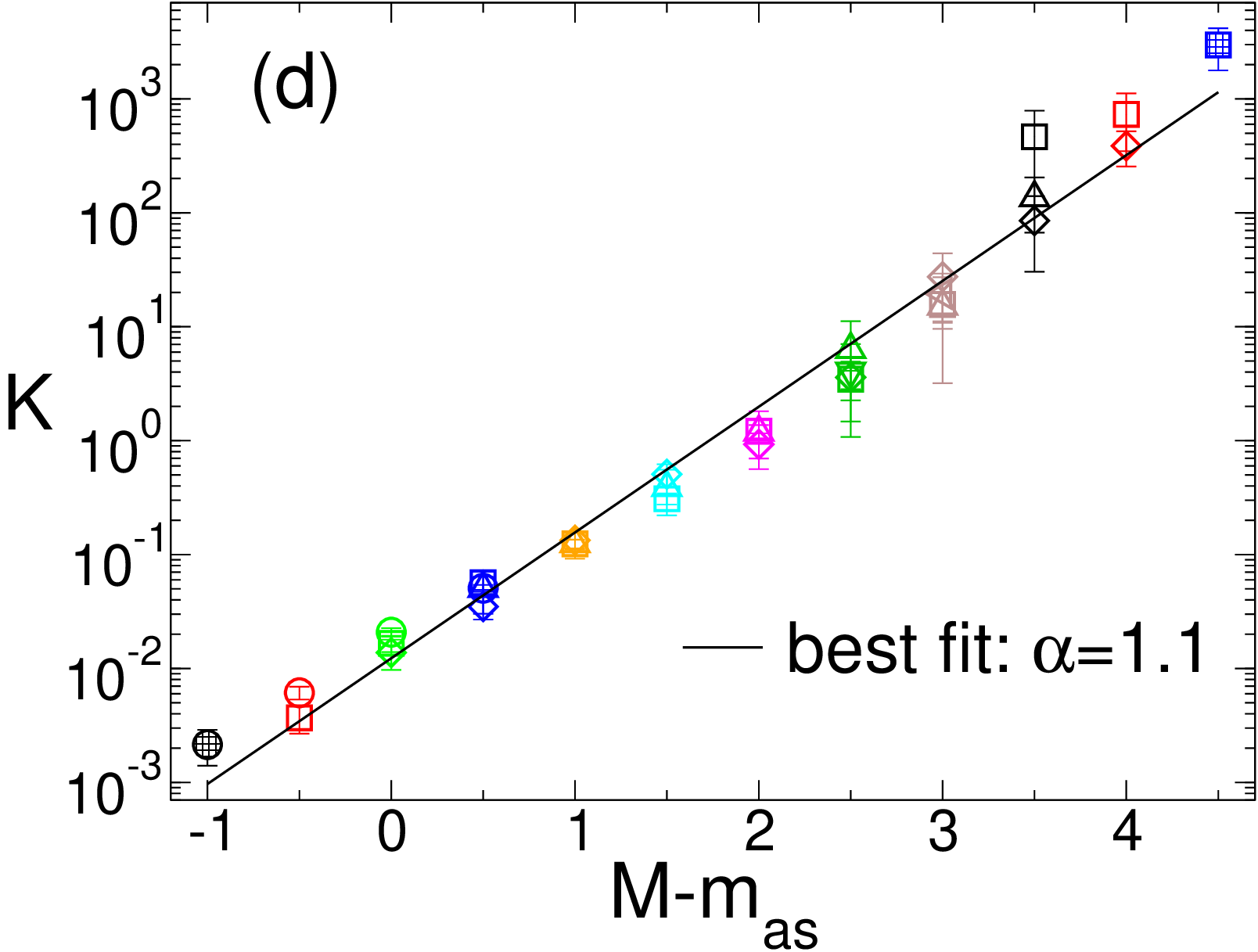}
\caption{\label{fig:fit}(a) Estimates of $b_{\rm{as}}$ as a function of the lower magnitude threshold $m_{\rm{th}}$ for different main shock ranges. (b-d) Estimated parameters of the Omori-Utsu relation~\eqref{eq:ou} for the conditional rates shown in Fig.~\ref{fig:ou}(a).}
\end{center}
\end{figure}

\paragraph*{Scaling function:}
Further support for the proposed self-similar OU relation comes from fitting the conditional triggering rates in Fig.~\ref{fig:ou} to Eq.~\eqref{eq:ss_ou} using a standard maximum likelihood estimator (MLE) for $p$, $c$ and $K$~\cite{davidsen14}, with $\tau \equiv {c^p}/{K}$ as in Eq.~\eqref{eq:ou}. Specifically, this allows us to estimate $c_{\Dm}$ and $K_{\Dm}$ (and consequently $\tau_{\Dm}$) and their behavior in more quantitative way as well as to establish whether the form of the scaling function~\eqref{eq:ss_ou} with a constant $p$ is appropriate. The corresponding results for $p$ are shown in Fig.~\ref{fig:fit}(b). There are no significant differences in the estimates for fixed $\Delta m$, though there is an increasing trend in $p$ with $\Delta m$ for large $\Delta m$. This can be attributed to the fact that the MLE slightly overestimates $p$ for large $\Delta m$ due to a detection issue of triggered events at late times
\footnote{
As discussed in Ref.~\cite{moradpour14}, there is a natural detection problem: declustering methods display an underestimation of the triggering rates at times later than about one year. The deviations from a power-law decay in the rates for $t>10^4$ hours are visible in Fig.~\ref{fig:ou}(a) and (c).}.
Indeed, an inspection of the rescaled rates (Fig.~\ref{fig:ou}(d)) confirms that for large $\Delta m$ the decay $t^{-p}$ is well-described by $p\approx 1.15$. This is also compatible with the values of $p$ reported in Fig.~\ref{fig:fit}(b) for smaller values of $\Delta m$, which suffer much less from the detection issue due to the more extended range of their power-law decay and follow the proposed functional form~\eqref{eq:ss_ou} very well (see Fig.~\ref{fig:ou}(a) for an example).

Moreover, the direct estimates for $c$ and $K$ show a clear scaling with $\Dm$, see Fig.~\ref{fig:fit}(c) and (d). Not only are the estimates statistically indistinguishable in almost all cases for fixed $\Dm$ but the scaling exponent $g$ is also consistent with the value obtained from the rescaling analysis above: Best fits give $g = 0.66 \pm 0.04$ and $\alpha = 1.10 \pm 0.03$, respectively. Here, the error bars correspond to $95\%$ confidence intervals (however, the systematic uncertainties mentioned above lead to higher error bars).
The behavior of $K$ shows in particular that the generalized productivity relation~\eqref{eq:prod} holds. This is further supported by Fig.~\ref{fig:ou}(c): The triggering rates collapse for sufficiently large values of $t$ under appropriate rescaling with $\alpha \approx 1$.

\paragraph*{Scaling relations:}
The self-similar OU relation provides a consistent picture where several direct MLE estimates of scaling exponents ($ g $, $ z $, $p$, $b_{\rm{as}}$ and $\alpha$) match estimates from scaling relations~\eqref{eq:sr_1} and~\eqref{eq:sr_2} between the different exponents. Having estimated the values of $ g $, $ z $ and $\alpha$ fully determines $p$ and $b_{\rm{as}}$. This gives $b_{\rm{as}} \approx 0.90$ and $p \approx 1.15$. 
Thus, the value of $p$ is consistent with the directly observed one. This is also true for $b_{\rm{as}}$. Fig.~\ref{fig:fit}(a) shows MLE estimates of $b_{\rm{as}}$~\cite{naylor09} as a function of the lower magnitude threshold $m_{\rm{th}}$ for different main shock ranges:  they yield $b_{\rm{as}} \approx 0.90$, which is clearly consistent with the value of $b_{\rm{as}}$ obtained from the scaling arguments. This value emerges independently of $m_{\rm{th}} \geq 2.5$.

Deviations from $b_{\rm{as}} \approx 0.90$, for main shocks with $M<3.5$, 
are consistent with the established presence of earthquake swarms in Southern California~\cite{vidale06}.
Swarms are typically associated with very specific geological settings and triggering mechanisms and 
are phenomena dominated by small magnitude events and characterized by larger $b_{\rm{as}}$-values.
Importantly, these higher $b_{\rm{as}}$ for small main shocks do not significantly affect the quality of the scaling collapse in Fig.~\ref{fig:ou}(b), where only four curves have $M<3.5$. 
Yet, we expect that swarm activity will lead to deviations from the self-similar OU relation for smaller magnitudes if not excluded from the triggering analysis.

\subsection{Physical origins vs observational limitations \label{sec:incom}}

Short-term aftershock incompleteness (STAI) is intrinsic to many earthquake catalogs. STAI arises from overlapping wave forms and/or detector saturation, in particular after large shocks, such that events are missed in the coda of preceding ones~\cite{kagan04,helmstetter06,lengline12}. This detection problem is not specific to earthquakes but a general problem related to crackling noise and the identification of ``events'' from recordings. One important consequence of STAI is an increase in the local magnitude of completeness~\cite{hainzl16} and, hence, an overestimation of the $c$-value in Eq.~\eqref{eq:ou} for large events~\cite{peng09,lengline12}. Thus, variations in $c$ with $\Dm$ for large $\Dm$ are typically expected due to STAI~\cite{kagan04,helmstetter06}.

Let us discuss several points that exclude STAI as a source of the self-similarity in the underlying triggering process observed for the data from SC.
First, STAI is not relevant for foreshocks ($\Dm<0$) and it has only minor effects for large aftershocks. Yet, the same scaling emerges over the whole range, namely for $-1 \leq \Dm \leq 5$, see Fig.~\ref{fig:ou}. Moreover, the scaling collapse shown in Fig.~\ref{fig:ou}(b) does even improve if we exclude rates with large $\Dm$: The rates at small times for large $\Dm$ are systematically smaller than what the scaling collapse of the other rates suggests.
Similarly, the estimated $c$-values are also systematically higher for the largest $\Dm$'s (see Fig.~\ref{fig:fit}(d)). Both effects are consistent with STAI. Thus, STAI is present, it leads to minor deviations from the proposed self-similar OU, but it cannot explain the observed self-similar behavior itself.

Second, the direct estimates of $\alpha$, $p$ and $b_{\rm{as}}$ are not significantly affected by STAI, since they either reflect the behavior at later times ($\alpha$, $p$) or do not vary with magnitude threshold $m_{\rm{th}}$ ($b_{\rm{as}} \approx 0.9$, see Fig.~\ref{fig:fit}(a)). With these values and the scaling relations~\eqref{eq:sr_1} and~\eqref{eq:sr_2}, the other two exponents $g$ and $z$ are fully determined.
Hence, the scaling of $c_{\Dm}$ with $\Dm$ is needed for consistency of the empirical picture where the complete set of five exponents is redundantly determined by direct estimates and scaling relations. One should thus understand the physical mechanisms generating non-trivial $c_{\Dm}$~\cite{nanjo07,narteau09,holschneider12}.

\begin{figure}[!t]
\begin{center}
\includegraphics[width = \columnwidth]{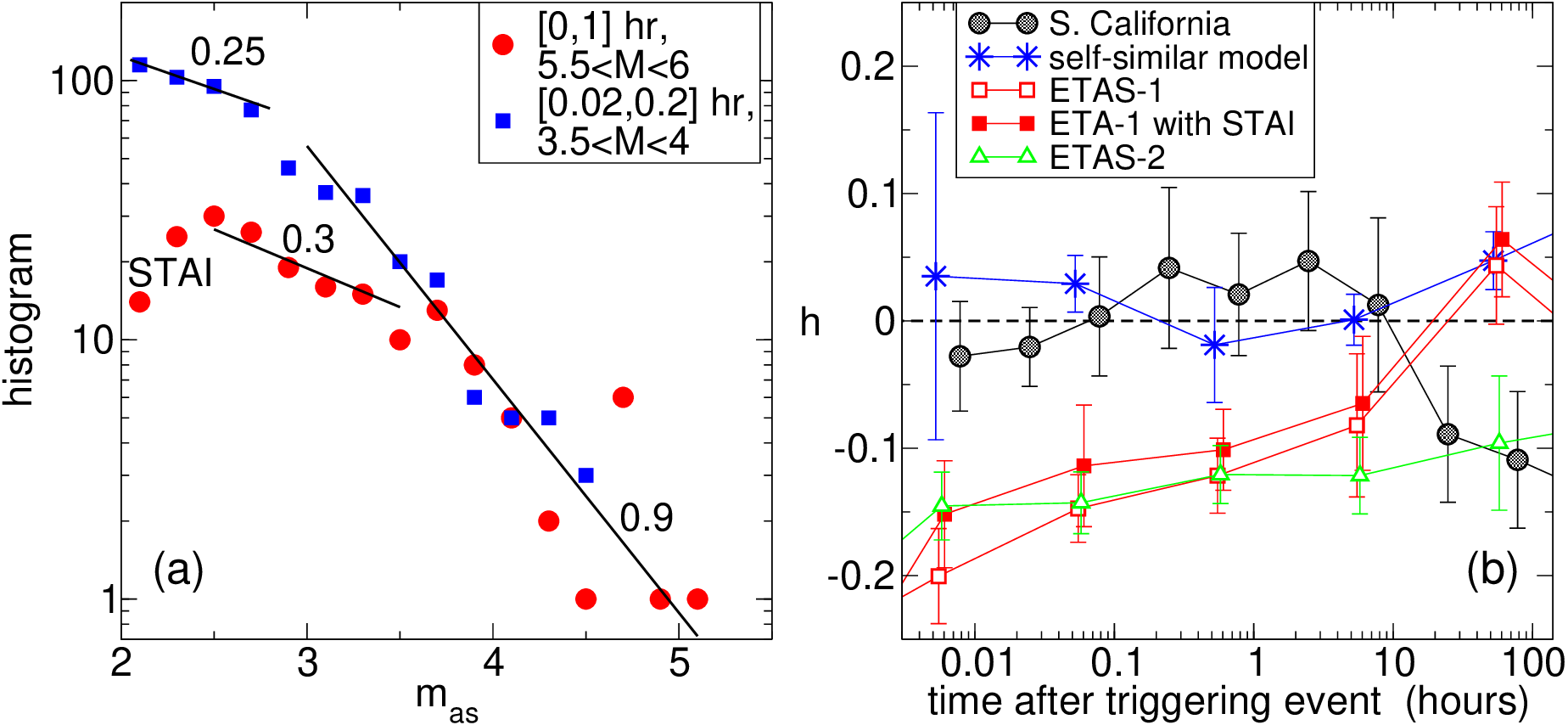}
\caption{\label{fig:3} (a) Stacked time-limited frequency-magnitude distributions for aftershocks. Different regimes are visible consistent with STAI and the proposed self-similar generalization of the OU relation. (b) Indicator of departure from self-similarity as a function of time, for the observed catalog and for four synthetic model catalogs. Only the self-similar model is compatible with the null indicator observed within one day from main shocks for the observed data. 
}
\end{center}
\end{figure}

Third, the limited effect of STAI is also evident from the number of triggered events over finite time intervals, defined in Eq.~\eqref{eq:ss_n_tr}. Fig.~\ref{fig:3}(a) shows two examples. For main shocks with $5.5 \leq M \leq 6.0$ and considering only aftershocks up to time $t^*=1$h, we observe the two regimes predicted by the proposed self-similar generalization of the OU relation: One at low $m$ consistent with the value of $z \simeq 0.24$ determined above, and a second regime at larger $m$ with an exponent consistent with $b_{\rm as} \simeq 0.9$. The typical effect of STAI of a temporary increase in the local magnitude of completeness is visible at the lowest magnitudes, where it leads to missing events and strong deviations from the proposed scaling behavior for aftershocks with $m_{\rm as} \leq 2.5$. 
This effect is not visible for smaller main shocks as the second example in Fig.~\ref{fig:3}(a) shows: For $3.5 \leq M \leq 4.0$, no deviations from the behavior predicted by the self-similar OU relation are visible on time scales longer than about 70sec.

The prediction by the self-similar OU relation of two power-laws in the time-limited frequency-magnitude distribution plus a third regime due to STAI as shown in Fig.~\ref{fig:3}(a) also allows us to revisit previously published work from a new perspective. In fact, it might be possible to partially connect documented temporal variations of the $b$-value~\cite{hainzl16} with (previously unnoticed) behavior in the time-limited frequency-magnitude distribution. 
Ref.~\cite{peng07} provides a specific example using a high-resolution catalog that has very carefully addressed the issue of STAI: The time-limited frequency-magnitude distribution of aftershocks in Japan shown in their Fig.~8b provides evidence for two different regimes above the magnitude of completeness. This supports the proposed self-similar OU relation beyond the catalog studied here, while fully taking STAI into account.

\subsection{Model simulation of the self-similar OU relation \label{sec:model}}

As a final step to confirm that our analysis is able to distinguish the proposed self-similar OU relation from non-self-similar versions for the conditional rates, we repeat it on synthetic catalogs generated by (a) the standard epidemic-type aftershock sequence (ETAS) model (which is not self-similar if the unconditional rates follow the classic OU relation with constant $c$ and $\alpha \neq b_{\rm{as}}$)~\cite{ogata88,vere-jones05,moradpour14}, and (b) our own variant built on the self-similar structure of Eq.~\eqref{eq:ss_ou} (see the Supplementary Information for more details).
If the conditional rates are not self-similar, there remains a dependence on the main shock magnitude $M$ that should be detectable in plots like that shown in Fig.~\ref{fig:ou}. By looking at groups of ETAS conditional rates with the same $\Dm$ and increasing $M$'s, we observe systematic decreases  (see the Supplementary Information).
The following estimator quantifies the trend with $M$: for each time $t$ and each $\Dm$ we have performed a linear fit of the log-log trend of $r(M-\Dm,t|M,0)$ vs $M$. This yields a slope $h_\Dm(t) \equiv {\rm Cov}(\log r,\log M)/{\rm Var(\log M)}$, which is supposed to be the same for all $\Dm$'s and all $t$'s. Hence the average slope $h(t) \equiv \sum_\Dm h_\Dm(t)  / \sum_\Dm 1$ is an indicator of the departure from self-similarity in the data.
In Fig.~\ref{fig:3}(b) we can see that, both for the data from Southern California and for the self-similar model, 
the indicator $h \approx 0$ 
 in the range $t\lesssim 1$ day, while at later times the statistics is not sufficient to perform its reliable estimate. For the ETAS models, the expected value $h = \alpha - b \approx -0.2$ is fairly well detected as well. 
We also find that self-similar behavior in the ETAS model cannot typically be induced by STAI: if STAI is introduced via one of the typical relations~\cite{helmstetter06}, our procedure continues to detect the absence of self-similarity $h \ne 0$ (Fig.~\ref{fig:3}(b)). See also the additional analysis in the Supplementary Information.
Thus, the comparison with synthetic data shows that our analysis scheme is sensible enough to detect self-similar behavior in the conditional OU rates, corroborating that the conditional triggering rates in Southern California are indeed self-similar.

\section{Conclusions \label{sec:concl}}

In summary, we have described a natural generalization of the OU relation that embodies self-similarity for event-event triggering in crackling noise. From a conceptual point of view, this provides some closure and an important piece in our understanding of event-event triggering where energy, distance and time appear in several scale-free empirical relations, either singularly (e.g.~the GR relation) or combined together. It has also profound consequences for probabilistic forecasting of aftershocks as well as modeling, as it implies that synthetic catalogs of relevant examples of crackling noise, including earthquakes, should be generated with algorithms reproducing the observed self-similarity. An important challenge for the future is to understand its physical origin, possibly with the help of lab experiments~\cite{davidsen05sg,baro13,maekinen15}.


\section{Methods}

\subsection{Catalog}
To test the validity of the self-similar OU relation, we analyze the relocated high-resolution Southern California catalog~\cite{hauksson12}. We consider all local earthquakes with magnitude $m\ge 2$ from 1982 to 2011 (101991 events).

\subsection{Aftershock identification}
We identify triggering relations between events and define aftershocks using the established methodology described in Refs.~\cite{baiesi04,zaliapin08,gu13amj,zaliapin13a,moradpour14}. The method quantifies the correlation between an event $i$ and a following event $j$ via the expression
\begin{equation}\label{eq:method}
	n(i,j) = C |t_i - t_j| |{\vec r}_i - {\vec r}_j|^{D_f} 10^{-b m_i},
\end{equation}
which estimates the expected number of events in the space-time window spanned by $i$ and $j$ with magnitude larger or equal to $m_i$. Here $t_i$ denotes the time of occurrence of event $i$, and $m_i$ its magnitude. $b=1.05$ is the estimated $b$-value for the full catalog. As in~\cite{gu13amj}, we use hypocenters ${\vec r}_i$ and the parameter $D_f=2.3$ for the fractal dimension and set $C=1$ without loss of generality. This leads to a threshold value $\log{n^*}=10.0$ for the identification of triggered events, i.e., only events with $n(i,j)< n^*$ are considered as plausible main shock-aftershock pairs. Among all events $i$ preceding $j$, the identification of the (most likely) trigger of $j$ results from selecting that with the lowest $n(i,j)$. Further details on the methodology can be found in Ref.~\cite{gu13amj}.

\subsection{Triggering rates}
For our analysis of triggering rates, we focus only on directly triggered events, namely we do not consider 
aftershocks of aftershocks. To obtain sufficient statistics, rates for all events in a given small magnitude range are stacked and averaged. In particular, triggered events with magnitude $M - m_{\rm{as}} \in [\Delta m,\Delta m+0.5)$ are selected for each trigger having a magnitude $M$ in the range $[M,M+0.5)$. 

\subsection*{}


\begin{acknowledgments}
J.D.~was financially supported by NSERC and the Alexander von Humboldt-Foundation. J.D.~would like to thank INFN and the University of Padova for their hospitality and support. We thank R. Shcherbakov for bringing Ref.~\cite{shcherbakov15} to our attention.
\end{acknowledgments}

\clearpage

\pagebreak
\widetext
\begin{center}
\textbf{\large Supporting Information for:\\Self-similar aftershock rates}
\end{center}
\setcounter{equation}{0}
\setcounter{section}{0}
\setcounter{figure}{0}
\setcounter{table}{0}
\setcounter{page}{1}
\makeatletter
\renewcommand{\thesection}{S\arabic{section}}
\renewcommand{\theequation}{S\arabic{equation}}
\renewcommand{\thefigure}{S\arabic{figure}}
\renewcommand{\bibnumfmt}[1]{[S#1]}
\renewcommand{\citenumfont}[1]{S#1}


We present a brief summary on how the synthetic catalogs analyzed in the main paper were generated and discuss their different statistical properties. We also show a direct comparison of the standard Omori-Utsu relation and the self-similar Omori-Utsu relation for the integrated aftershock rates. Enlarged views of some of the figures in the main text are also presented for better readability.

\onecolumngrid


\section{Standard ``Epidemic Type Aftershock Sequence'' (ETAS) models}

In the ETAS model any event can trigger other events (event-event triggering) and the total rate of event occurrence with magnitudes $m$, at time $t$ and at location $\vec{r}$ is defined as~\cite{kagan81,ogata88,ogata06,moradpour14}:
	\begin{equation}\label{eq::total_rate}
		\lambda(t,\vec{r})= \mu_b(\vec{r}) +  \sum_{t_i<t} \phi_{m_i}(\vec{r}-\vec{r}_i, t-t_i).
	\end{equation} 
$\mu_b(\vec{r})$ corresponds to the rate of occurrence of background events, which can vary in space. Each of the background events $j$ can be a main shock and trigger events (aftershocks) with a rate given by $\phi_{m_j}(\vec{r}-\vec{r}_j, t-t_j)$, corresponding to a spatially non-homogeneous and time-varying marked Poisson process. These aftershocks can also trigger other events and then we can have a cascade of events as described by Eq.~\eqref{eq::total_rate}. Here, the magnitude distribution of both background events and triggered events is assumed to have the same functional form and to be independent of the past activity. Specifically, the magnitude of events are chosen according to the normalized GR probability distribution
	\begin{equation}\label{eq::GR_distribution}
		P_m(m)=b \; \ln{10} \; 10^{-b(m-m_c)},
	\end{equation}	
where $m_c$ is the lower magnitude threshold of events, selected in the model used to generate the surrogate catalog. 

As follows from the second term of Eq.~\eqref{eq::total_rate}, the model assumes that the triggering processes lead to a simple linear superposition in terms of the rates. These rates, $\phi_{m_i}(\vec{r}-\vec{r}_i, t-t_i)$, quantify the spatiotemporal distribution of triggered events at spatial distance $\vec{r}-\vec{r}_i$ and temporal distance $t-t_i$ from a trigger (main shock) with main shock magnitude $m_i$. Typically, the functional form is assumed to factorize into three terms:
	\begin{equation}
		\phi_{m_i}(\vec{r},t)=\rho(m_i) \; \psi(t) \; \zeta_{m_i}(\vec{r}).
                \label{eq:fac}
	\end{equation}
Here, $\rho(m_i)$ is the number of events triggered by the event $i$, which is assumed to follow a Poissonian distribution with an average of $\langle\rho(m_i)\rangle$. The latter follows the standard productivity relation:
	\begin{equation}
		\langle\rho(m_i)\rangle=K \; 10^{\alpha_\text{ETAS}(m_i-m_c)}, \label{eq::productivity_law}
	\end{equation}
where $m_c$ is again the lower magnitude threshold.

$\psi(t)$ in Eq.~\eqref{eq:fac} is the normalized temporal distribution of triggered events at time $t$ after the main shock. It is assumed to follow the standard Omori-Utsu relation:
\begin{equation}\label{eq::Omori_law}
	\psi(t)=\dfrac{(p-1) \; c^{p-1}}{(t+c)^p}.
\end{equation}

$\zeta_{m_i}(\vec{r})$ in Eq.~\eqref{eq:fac} is the normalized spatial distribution of triggered events with distance $\vec{r}$ from the main shock. The direction of this distance vector is typically chosen at random. The distribution of its length, $r$, corresponds to $P_{m_i}(r) \equiv \int_0^{2\pi} \zeta_{m_i}(\vec{r}) r d\phi $ and for earthquakes it is given by~\cite{moradpour14,hainzl15}
\begin{equation} \label{eq::spatial_distribution}
		P_{m_i}(r) =\begin{cases}
		A_0\dfrac{q\, r^{\gamma}}{L_{m_i}^{\gamma+1}} 
\left(\dfrac{r^{\gamma+1}}{L_{m_i}^{\gamma+1}}+1\right)^{-(1+\frac{q}{\gamma+1})} & \mbox{if  $r < 10$km}, \\[1cm] 
		A_1\dfrac{d\, r^{\gamma}}{L_{m_i}^{\gamma+1}}
\left(\dfrac{r^{\gamma+1}}{L_{m_i}^{\gamma+1}}+1\right)^{-(1+\frac{d}{\gamma+1})}
& \mbox{if $r > 10$km} .
		\end{cases}
\end{equation}
In this equation $A_0$ and $A_1$ are normalization factors and $\gamma$, $q$ and $d$ are constants describing the functional behavior over different ranges. $L_{m_i} = L_R/2$, where the rupture length, $L_R$, scales with the magnitude of the event, $m_i$, according to~\cite{wells94,kagan02,leonard10,gu13amj} as:
	\begin{equation}
		L_R= l_0 \times 10^{\sigma m_i},
                \label{eq::rupture_length}
	\end{equation}
where $l_0$ and $\sigma$ are constants with typical values of $l_0$ in the range of $10$ -- $20$m and $0.4 \leq \sigma \leq 0.5$. It is important to realize that $\vec{r}$ in Eq.~\eqref{eq:fac} or equivalently $\vec{r}_i$ in Eq.~\eqref{eq::total_rate} can correspond to different quantities~\cite{moradpour14}. For example, $\vec{r}$ can be the epicenter-to-epicenter distance or it can be the epicenter-to-source distance. In the latter case, $\vec{r}_i$ is the location of a randomly chosen point along the main shock rupture corresponding to the source. It is typically assumed that the main shock rupture is centered at the epicenter of the main shock and has a random orientation. For a given main shock, the main shock rupture is kept fixed and the source varies for each aftershock.

The first ETAS catalog we analyze here (``ETAS-1'') corresponds to such an anisotropic case and mimics the seismic activity in Southern California (including a spatially varying background rate) as closely as possible. It was first discussed in Ref.~\cite{moradpour14}, where it was labeled model VII. There is also a version that includes the effects of short-term aftershock incompleteness (STAI, model I in Ref.~\cite{moradpour14}) which we call ``ETAS-1 with STAI'' and consider here as well  --- the specific details on how STAI is modeled are described below in the section ``Short-term aftershock incompleteness''. The most relevant specific parameter values of the ETAS model are:
$b = 1.09$, $\alpha_\text{ETAS} = 0.88$, $p = 1.1$, $K=0.1735$, $c=0.0001$days, $m_c=2.5$. For all other parameters, see Tables I and II in Ref.~\cite{moradpour14}. Note that model VII has 51447 events while model I has 37092 events. As in Ref.~\cite{moradpour14}, we use $D_f=1.6$ as well as $b=1.01$, $\log{n^*}=7.0$ and $b=1.09$, $\log{n^*}=6.0$ for model I and model VII, respectively, for identifying triggered events.

A further, somewhat simpler ETAS catalog (``ETAS-2''), generated in a $600\times 600$ km area, is also analyzed for a direct comparison with the self-similar model and to test whether some features such as the boundary conditions yield spurious effects or not. It is generated with open boundary conditions (events outside of the square area are forgotten), with a homogeneous background rate (background events have epicenters that are at a distance of at least $60$ km from the border of the seismic area) and isotropic spatial kernel (epicenter-epicenter distances) with parameters $\sigma=0.45$, $q=d=0.6$, $\gamma=1$, $l_0=20$m.
Other parameters are  $b = 1.08$, $\alpha_\text{ETAS} = 0.88$, $p = 1.1$, $K=0.18$, $c=10$sec, $m_c=2$.
The catalog is composed of $121370$ events generated over $30$ years starting from  the background activity of two events per day. 
To identify triggering relations and aftershocks using the same methodology as in the other cases, we choose $D_f=2.0$, $b=1.08$, $\log{n^*}=9.5$.

With the same procedure described for ETAS-2, we also generated a catalog ETAS-3 with STAI using Eq.~\eqref{eq:STAI}. In this case, we have a constant $c=16$sec and $\alpha_\text{ETAS}=b=1$, i.e., it is the a special case of an ETAS model with self-similarity if STAI were not present. Besides $K=0.16$, the other parameters coincide with those of ETAS-2. The final catalog consists of $83700$ events.
To identify triggering relations and aftershocks using the same methodology as in all other cases, we choose $D_f=2.0$, $b=1.0$, $\log{n^*}=9.5$.


\section{Self-similar aftershock model}

For the self-similar aftershock model, Eq.~\eqref{eq:fac} is replaced by the following factorization
	\begin{equation}
		\phi_{m_i,m}(\vec{r},t)=\rho(m_i) \; \psi_{m_i,m}(t) \; \zeta_{m_i}(\vec{r}),
                \label{eq:fac_ss}
	\end{equation}
such that there is now an explicit dependence on $m$. 
While $\zeta_{m_i}(\vec{r})$ is as in the ETAS model, this is not true for the other two terms. Specifically, $\langle\rho(m_i)\rangle$ is assumed to follow the generalized productivity relation:
	\begin{equation}
		\langle\rho(m_i)\rangle=K \; 10^{b_{\rm as}(m_i-m_c)}, \label{eq::productivity_law_ss}
	\end{equation}
with $K=\frac{c_0}{b_{\rm as} \tau_0 (p-1) \ln{10}}$, $c_0$ and $\tau_0$ constants to be defined precisely below and $m_c$ is again the lower magnitude threshold.

Moreover, $\psi_{m_i,m}(t)$ in Eq.~\eqref{eq:fac_ss} is assumed to follow the proposed self-similar generalization of the Omori-Utsu relation:
\begin{equation}\label{eq::Omori_law_ss}
	\psi_{m_i,m}(t)=\dfrac{(p-1) \; c_{\Delta m}^{p-1}}{(t+c_{\Delta m})^p}.
\end{equation}
The self-similarity is captured by the fact that $c$ is now a function of the magnitude of the triggered event, $m$, and of the magnitude of the trigger, $m_i$, such that $c=c_0 10^{g (m_i-m)} \equiv c_{\Delta m}$ only depends on the magnitude difference $\Dm=m_i-m$. Note that combining Eqs.~\eqref{eq::productivity_law_ss} and~\eqref{eq::Omori_law_ss}, we recover the proposed self-similar generalization of the Omori-Utsu relation:
\begin{equation}\label{eq::ss_Omori_law}
	r(m,t|m_i,0)=\dfrac{1}{\tau_\Dm (t/c_\Dm+1)^p},
\end{equation}
with $\tau_\Dm=\tau_0 10^{-z \Dm}$ and 
$z=b_{\rm as} -g$
for consistency.

To obtain synthetic catalogs based on this model, we first generate a list of background events and spread aftershocks around them following the $\zeta_{m_i}(\vec{r})$ distribution. 
Since the rate of events with magnitude $m$ after each event with magnitude $m_i$ depends on $\Delta m = m_i-m$, in order to generate the list of aftershocks we discretize magnitudes with a small resolution $\epsilon=0.01$, and for each interval $[m,m+\epsilon)$ in the range $2\le m < 9$ we compute the average number of expected aftershocks
\begin{equation}
n_\epsilon(m|m_i,0)  = \epsilon \int_0^\infty r(m,t|m_i,0) d t
\end{equation}
with conditional rates $r$ given in Eq.~\eqref{eq::ss_Omori_law}.
The actual number of aftershocks in this magnitude interval is then drawn from a Poissonian distribution with this average, and for each of these aftershocks the time from the main shock is drawn from the distribution specified by Eq.~\eqref{eq::Omori_law_ss}. All events generated in this way may then give rise to their own aftershocks, and the iteration of this procedure ends when no event within the time span of the catalog remains to be processed. A final time-ranking yields the catalog. Note that the aftershocks GR relation with $b_{\rm as} = z+g$
is implicitly generated through the previous procedure.

For generating synthetic catalogs with the self-similar aftershock model, we use the estimated values for Southern California (see main paper) to fix the parameters:  
$p = 1.15$, $g = 0.66$, $z = 0.24$, $c_0=210$ sec, $\tau_0=10^4$ sec. All other parameters of the model as well as the parameters for the aftershock detection method are the same as for ETAS-2. The specific catalog we consider in the main text contains $112176$ events. 
Background events are drawn from a GR relation with $b=1.08 > b_{\rm as}=0.9$. Due to upper magnitude cut-off and the choice of the parameters, the branching ratio is less than one, making the process sub-critical~\cite{zhuang13}. Yet, the catalog turns out to be composed mostly of aftershocks ($\approx 82$\%).


\section{Comparison of ETAS and self-similar model}

\begin{figure*}[!h]
\begin{center}
\includegraphics[width = 8cm]{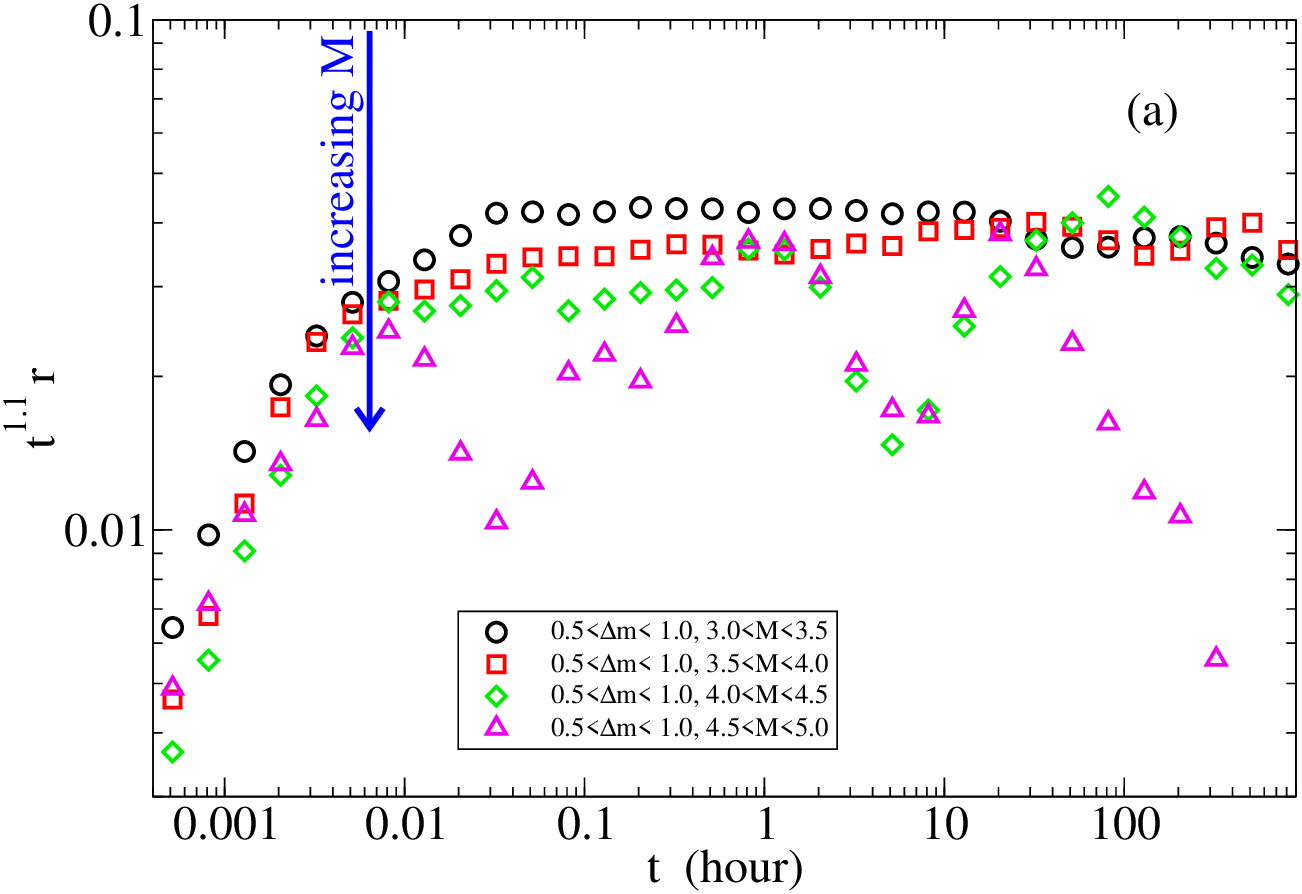}
\hskip 2mm
\includegraphics[width = 8cm]{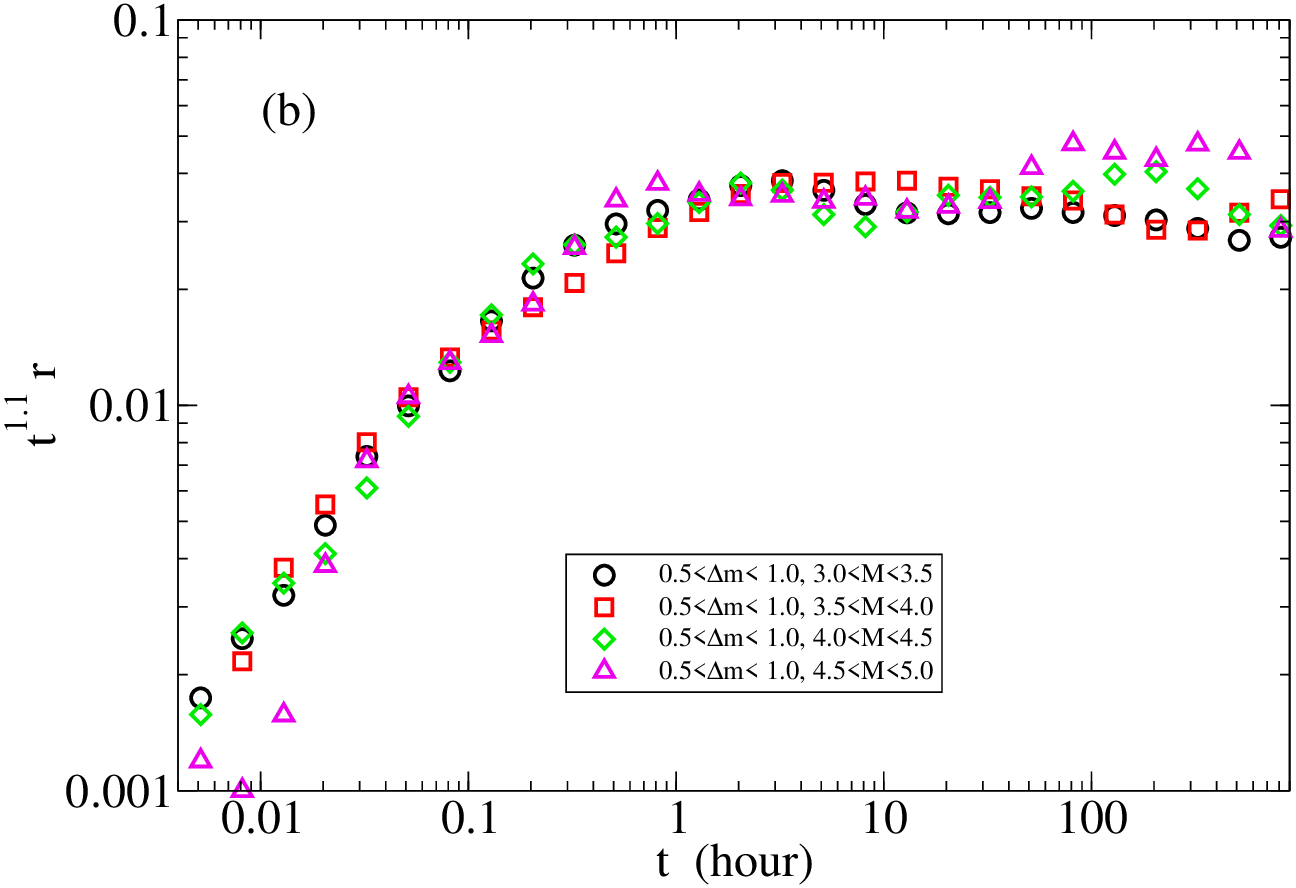}
\caption{(Color online) Examples of conditional rates rescaled with time according to $t^{1.1}$ of some aftershocks in a group with the same $\Delta m$ and different $M$'s,  generated in (a) the ETAS-2 catalog and (b) the catalog of the self-similar model, based on the scaling discussed in the main text.\label{fig:ex-ss}}
\end{center}
\end{figure*}

Catalogs without self-similarity (from ETAS models with $\alpha_\text{ETAS} \neq b$) display a trend in each group of conditional triggering rates with the same $M-m$.
For our specific choice of parameters, the rates decrease with increasing main shock magnitude, see an example in Fig.~\ref{fig:ex-ss}(a).
In the figure one can also see that the detection procedure is precise within about $10$--$24$ hours from the main shock, then it does not display a clear scaling anymore, partly due to poorer statistics in the tails of the decay.
This feature is to some degree also present in the catalog generated with the self-similar model, as shown in  Fig.~\ref{fig:ex-ss}(b).
More importantly, this figure clearly indicates that the self-similar behavior is detected, as there is no significant trend in the curves for increasing $M$.

\clearpage

\begin{figure}[!t]
\begin{center}
\floatbox[{\capbeside\thisfloatsetup{capbesideposition={left,top},capbesidewidth=5.cm}}]{figure}[\FBwidth]
{\caption{(Color online) Averaged conditional aftershock rates for different ranges of main shock and aftershock magnitudes for the ETAS-1 catalog with STAI as described by Eq.~(\ref{eq:STAI}).}\label{fig:STAI}}
{\includegraphics[width = 10.6cm]{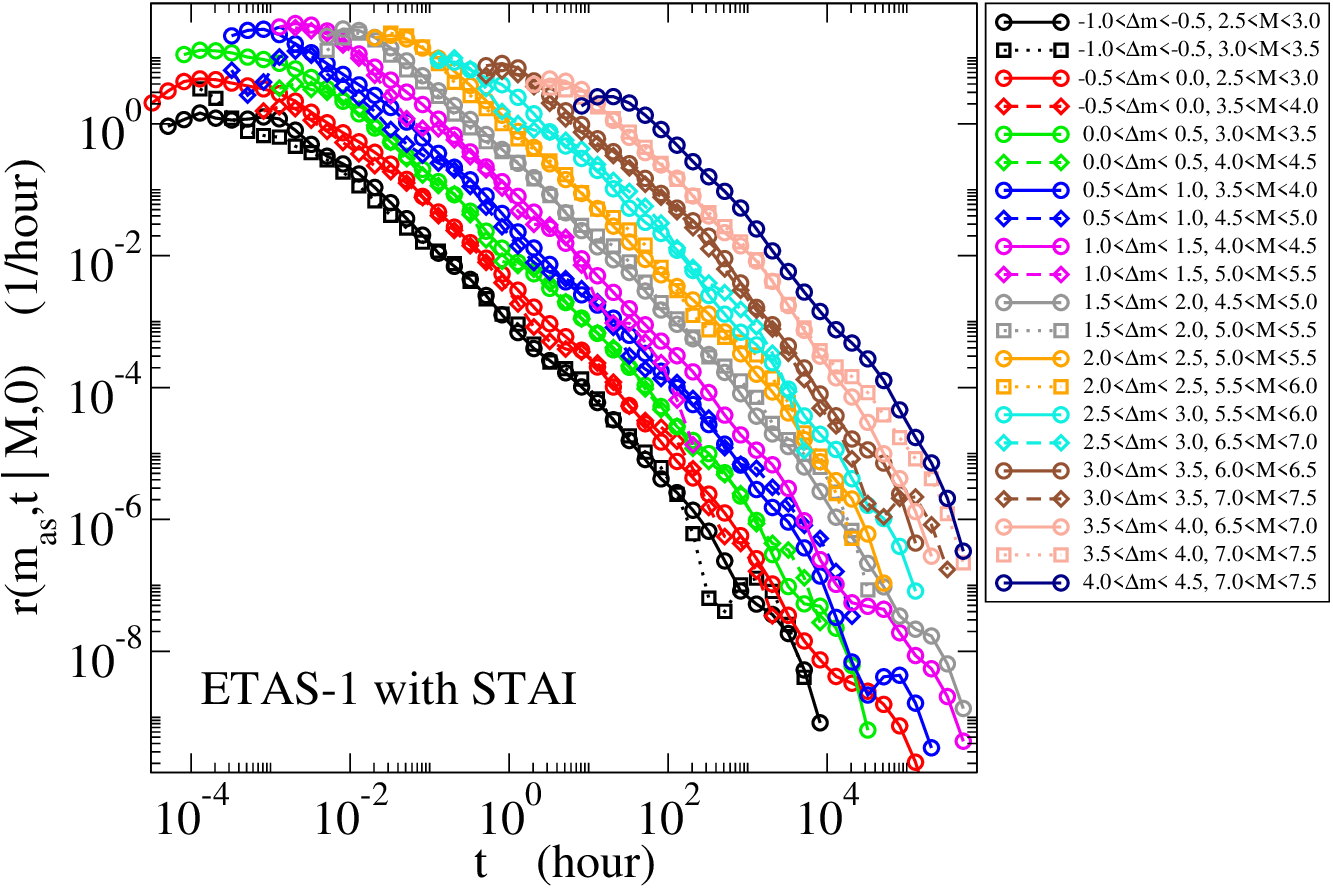}}

\vskip 3mm

\floatbox[{\capbeside\thisfloatsetup{capbesideposition={left,top},capbesidewidth=5.cm}}]{figure}[\FBwidth]
{\caption{(Color online) Averaged conditional aftershock rates for different ranges of main shock and aftershock magnitudes for the ETAS-3 catalog with STAI as described by Eq.~(\ref{eq:STAI}). }\label{fig:STAI2}}
{
\includegraphics[width = 10.6cm]{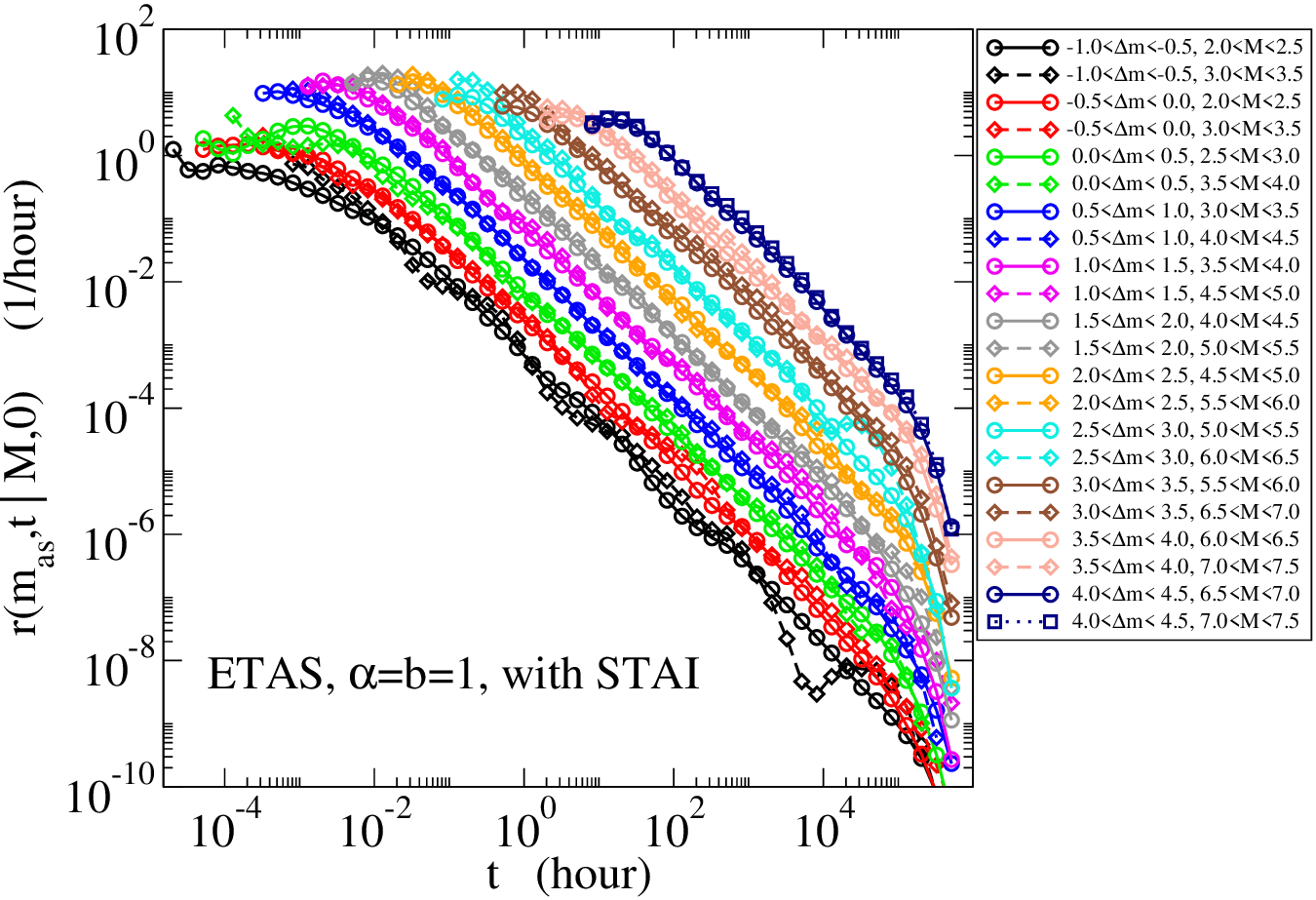}}

\vskip 4mm

\floatbox[{\capbeside\thisfloatsetup{capbesideposition={left,top},capbesidewidth=5.cm}}]{figure}[\FBwidth]
{\caption{(Color online) Averaged conditional aftershock rates for different ranges of main shock and aftershock magnitudes for the catalog generated by the self-similar model.
Contrary to the rates in Figs.~\ref{fig:STAI} and~\ref{fig:STAI2}, these conditional rates exhibit the same scaling as the data from Southern California reported in Fig.~1(a) of the main text.}\label{fig:our}}
{\includegraphics[width = 10.6cm]{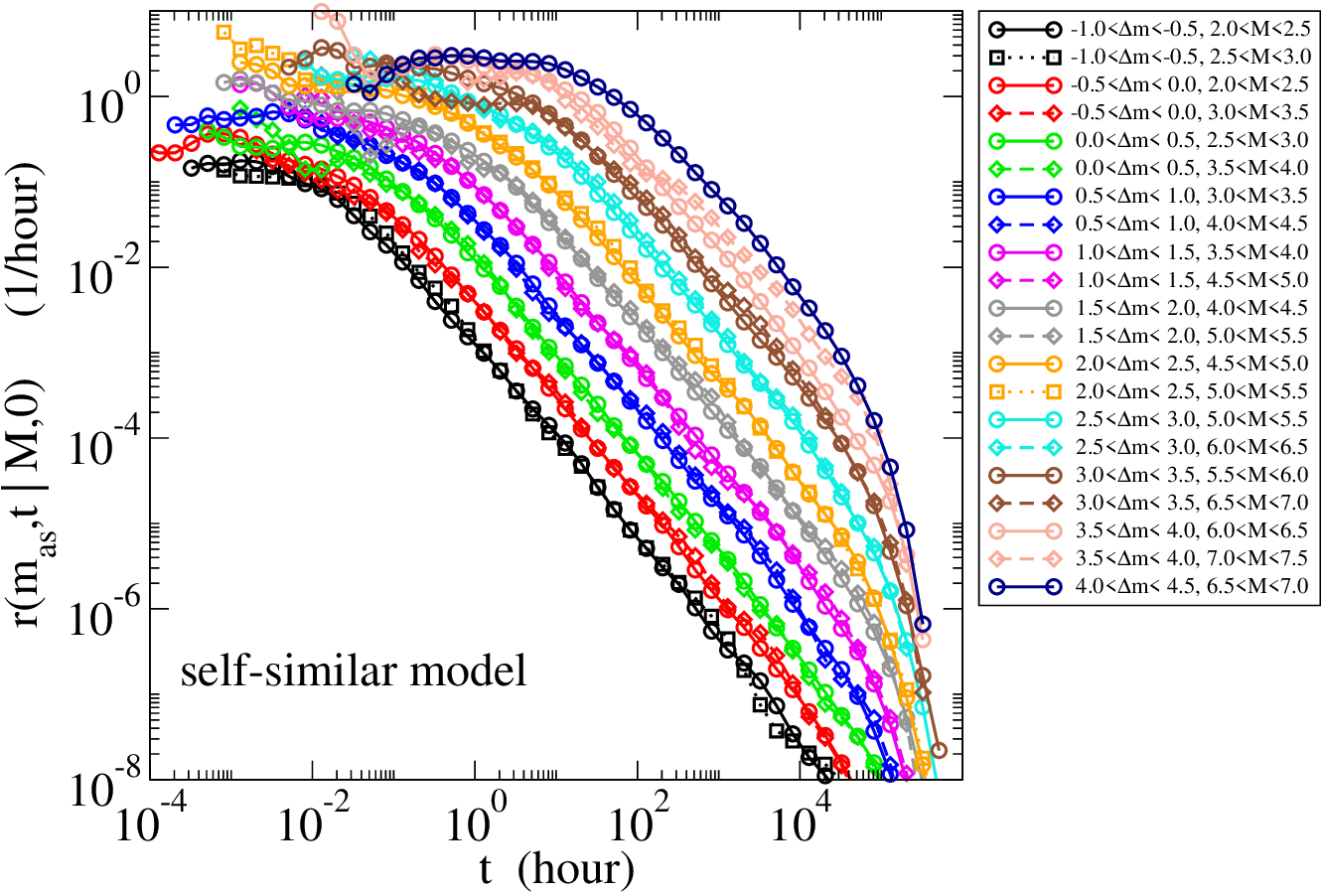}}
\end{center}
\end{figure}

\section{Short-term aftershock incompleteness}

As in many other studies (e.g., Refs~\cite{gu13amj,moradpour14}), the formula we adopt to mimic short-term aftershock incompleteness (STAI) in the ETAS catalogs follows the trend of STAI reported by~\citet{helmstetter06}.
Events after each event $i$ of magnitude $m_i=M$ are removed from the catalog if their magnitude is below a threshold value $m_{\rm det}(t,M)$ that depends both on $M$ and on the time lag $t$ after event $i$. This threshold is given by 
	\begin{equation}
\label{eq:STAI}
	m_{\rm det}(t,M) = M-4.5-0.75 \log_{10} t\,.
	\end{equation}
The visual inspection of the conditional Omori rates for the ETAS-1 catalog with STAI (Fig.~\ref{fig:STAI}) shows that they do not match the global form of scaling for Southern California visible in Fig.~1(a) of the main text: for increasing $\Delta m$, the maximum rate first increases and then decreases such that no scaling collapse is possible. This is also true for the ETAS-3 catalog (Fig.~\ref{fig:STAI2}) generated with the special condition $b=\alpha=1$ which yields self-similarity before including STAI. The absence of a scaling collapse is not surprising since STAI as described by Eq.~\eqref{eq:STAI} should not affect foreshock-main shock pairs and it should only minimally affect main shock-aftershock pairs with small magnitude differences.
In principle the above equation for STAI might entail also a form of self-similarity (as $t$ scales as $10^{4/3(M-m_{\rm det})}$) that could generate $c$ values artificially depending on $M-m$, similar to what we observe in South California. This is, however, not the case: Not only are there still significant variations independent of $\Delta m$ as follows from Fig.~3 of the main text but also Figs.~\ref{fig:STAI} and~\ref{fig:STAI2} show that there is no scaling behavior in $\tau_{\Dm}$. In contrast, the overall scaling in Fig.~1(a) is well reproduced by our self-similar model (Fig.~\ref{fig:our}).

\section{Conditional vs. integrated self-similar Omori-Utsu rates}
\begin{figure}[!h]
\begin{center}
\includegraphics[width = 8cm]{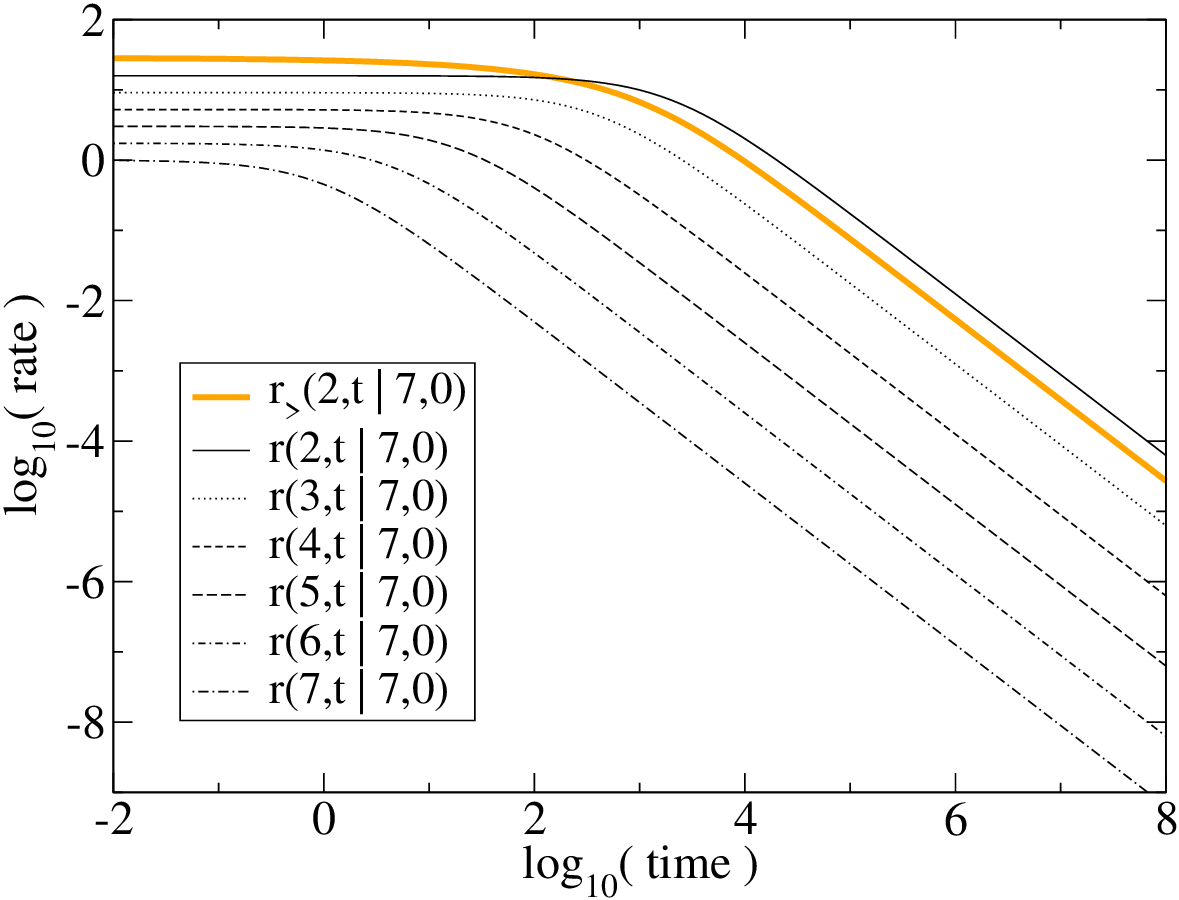}
\caption{\label{fig:comp}(Color online) Comparison between the conditional and the integrated self-similar Omori-Utsu relation for $M=7$, $m_\text{th}=2$, and several values of the aftershock magnitudes. 
In this example we set $c_0 = \tau_0 = 1$,
$g = 0.66$,
$z = 0.24$, and
$p = 1.15$.
}
\end{center}
\end{figure}

If the conditional aftershock decay rate $r(m,t|M,0)$ given by Eq.~(3) in the main text (equivalent to Eq.~\eqref{eq::ss_Omori_law}) is assumed, the integrated rate 
 $r_>(m_{\rm{th}},t|M,0) \equiv \int_{m_{\rm{th}}}^\infty  r(m,t|M,0)\, d m$ that arises,
\begin{eqnarray}
\label{hyp}
r_>(m_{\rm{th}},t|M,0) = \frac 1 {\tau_0 \ln 10} &&
\left[
\frac {10^{(M-m_{\rm th})z}} z {\rm\; _2F_1}\left(p,-\frac z g,1-\frac z g, -\frac t {c_0} {10^{-(M-m_{\rm th})g}}\right)
- \frac 1 z               {\rm\; _2F_1}\left(p,-\frac z g,1-\frac z g, -\frac t {c_0}\right)\right.\nonumber\\
&&\left.+ \frac{(t/c_0)^{-p}}{z+pg}      {\rm\; _2F_1}\left(p,p+\frac z g,1+p+\frac z g, -\frac {c_0} t\right)
\right]
\end{eqnarray}
includes hypergeometric functions
\begin{equation}
 {\rm\; _2F_1}(a,b,c,x)\equiv \sum_{k=0}^\infty\frac{(a)_k(b)_k}{(c)_k} \frac{x^k}{k!} 
\end{equation}
where $(a)_k \equiv a (a+1)(a+2)\cdots(a+k)$.
Despite the complicated form, the integrated rates described by (\ref{hyp}) have a a functional form very similar to the classical Omori-Utsu relation for realistic parameters as shown in Fig.~\ref{fig:comp}, the only difference being that the transition region between the constant regime at short times and the power-law decay at longer times is a little bit broader than in the classic Omori-Utsu relation. This difference vanishes for $g=0$. Most importantly, the integrated rates inherit the self-similarity with respect to $M-m_{\rm{th}}$: Magnitudes enter in (\ref{hyp}) only through the combination $M - m_{\rm th}$. A consequence is that, for example, the rate of aftershocks with magnitude $\ge 4$ of a main shock with $M=7$ is the same of aftershock with magnitude $\ge 2$ of a main shock with $M=5$, because $M - m_{\rm th}=3$ in both cases.

\section{Enlarged view}
\begin{figure*}[!t]
\begin{center}
\includegraphics[width = 13.cm]{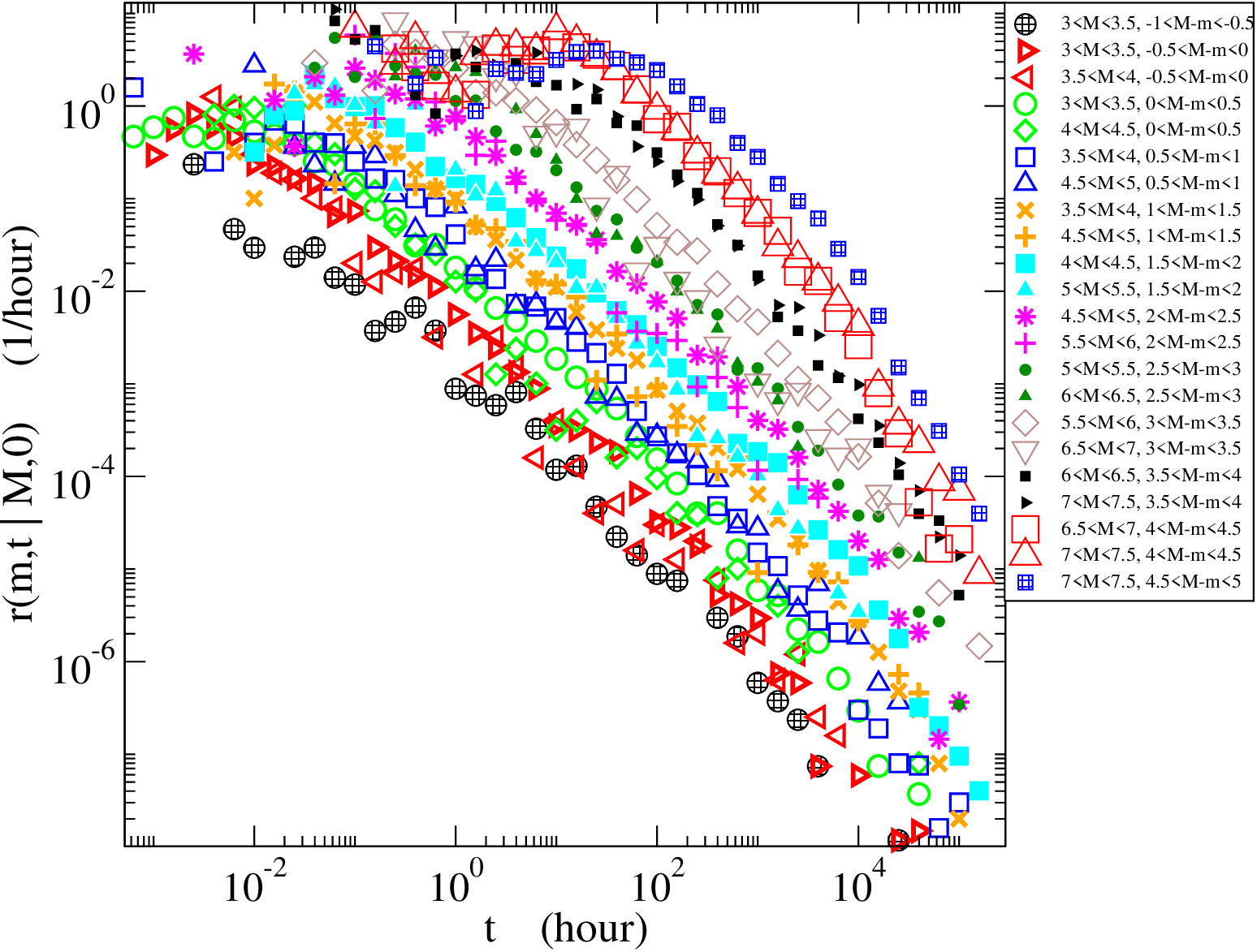}
\caption{(Color online) Enlargement of Fig.~1(a) of the main text, with less data sets.
Here $m$ denotes the aftershock magnitude and as usual $M$ the mains shock magnitude.
\label{fig:enlarged-a}
}
\end{center}
\end{figure*}

\begin{figure*}[!t]
\begin{center}
\includegraphics[width = 12.cm]{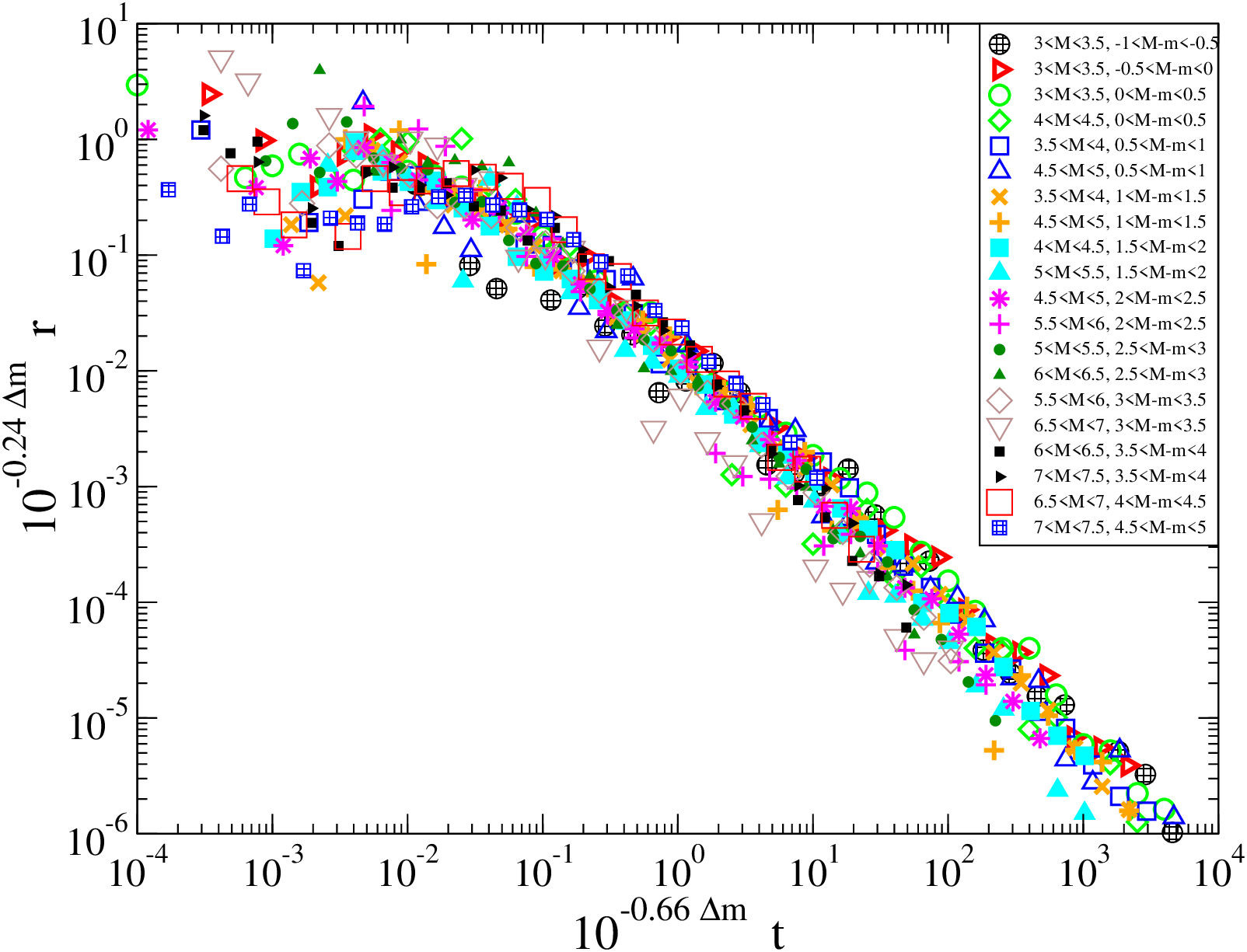}
\caption{(Color online) Enlargement of Fig.~1(b) of the main text, with less data sets.
Data collapse and no appreciable trend with $\Dm$ is visible.
\label{fig:enlarged-b}
}
\end{center}
\end{figure*}

For a better visualization, we enlarge Fig.~1(a) and Fig.~1(b) of the main text, see 
respectively Fig.~\ref{fig:enlarged-a} and Fig.~\ref{fig:enlarged-b}.
Also for the sake of a clearer visual understanding, in these figures we use a more contrasted graphics and
we (mostly) plot only the first and the eventual third set of data for each $\Dm$.
Fig.~\ref{fig:enlarged-b} helps appreciating the data collapse on a single scaling form.

\clearpage

\clearpage

%


\end{document}